\PassOptionsToPackage{table}{xcolor}
\documentclass[acmsmall,screen,natbib=false,nonacm]{acmart}

\usepackage{tabularx}
\usepackage{booktabs}
\usepackage{subcaption}

\usepackage{tikz}
\usetikzlibrary{calc}

\definecolor{periwinkle}{rgb}{0.50,0.50,1.0}

\definecolor{ChristmasTree}{rgb}{0.13,0.55,0.13}

\definecolor{asparagus}{rgb}{0.53, 0.66, 0.42}

\usepackage{ifluatex}
\ifluatex
  \usepackage[utf8]{luainputenc}
\else
  \usepackage[utf8]{inputenc}
\fi

\usepackage{xcolor}
\usepackage{tikz}
\usetikzlibrary{patterns}
\usepackage{colortbl}
\usepackage{pgfplots}
\usepackage{pgfplotstable}
\pgfplotsset{compat=1.17}
\usepackage{adjustbox}
\usepackage{balance}

\usepackage{pgfplots}
\usepackage{pgfplotstable}

\usepackage{graphicx}
\usepackage{color}
\usepackage{booktabs}
\usepackage{siunitx}
\usepackage{url}
\usepackage{nicefrac}
\usepackage{float}
\usepackage[mincitenames=1,maxcitenames=1,maxbibnames=1,backend=biber,giveninits=true,doi=true,isbn=false,url=true]{biblatex} %
\ifluatex
  \usepackage{fontspec}
  \setmathrm{TeX Gyre Termes}
  \setmathsf{TeX Gyre Heros}
  \setmathtt{Tex Gyre Cursor}
  \usepackage[english]{selnolig}
\fi

\usepackage{caption}

\usepackage{paralist}
\usepackage{tensor}
\usepackage{listings}
\input{cfdlang.sty}
\usepackage{csquotes}

\usepackage{comment}
\usepackage{enumitem}
\setcopyright{acmlicensed}
\acmJournal{TRETS}
\acmYear{2022} \acmVolume{1} \acmNumber{1} \acmArticle{1} \acmMonth{1} \acmPrice{15.00}\acmDOI{10.1145/3563553}

\usepackage{tikz}

\usepackage{cleveref}

\newcommand{\amd}{AMD~EPYC~7282}

\usepackage[acronym]{glossaries}
\glsdisablehyper
\newacronym{alap}{ALAP}{as-late-as-possible}
\newacronym{ast}{AST}{abstract syntax tree}
\newacronym{cu}{CU}{compute unit}
\newacronym{cfd}{CFD}{computational fluid dynamics}
\newacronym{cse}{CSE}{common subexpression elimination}
\newacronym{dce}{DCE}{dead code elimination}
\newacronym{dsl}{DSL}{domain-specific language}
\newacronym{dsp}{DSP}{digital signal processor}
\newacronym{ff}{FF}{flip-flop}
\newacronym{ffi}{FFI}{foreign function interface}
\newacronym{fifo}{FIFO}{first in - first out}
\newacronym{gemm}{GEMM}{generic matrix-matrix multiplication}
\newacronym{hbm}{HBM}{high-bandwidth memory}
\newacronym{hls}{HLS}{high-level synthesis}
\newacronym{hpc}{HPC}{high-performance computing}
\newacronym{ii}{II}{initiation interval}
\newacronym{ir}{IR}{intermediate representation}
\newacronym{isa}{ISA}{Instruction Set Architecture}
\newacronym{isl}{ISL}{integer set library}
\newacronym{lut}{LUT}{lookup table}
\newacronym{ml}{ML}{machine learning}
\newacronym{mlir}{MLIR}{multi-level intermediate representation}
\newacronym{pc}{PC}{pseudo-channel}
\newacronym{plm}{PLM}{private local memory}
\newacronym{qor}{QoR}{quality of result}
\newacronym{raw}{RAW}{read-after-write}
\newacronym{sem}{SEM}{spectral element methods}
\newacronym{sll}{SLL}{super long line}
\newacronym{slr}{SLR}{super logic region}
\newacronym{soc}{SoC}{system-on-chip}
\newacronym{tc}{TC}{TensorComprehensions}
\newacronym{teil}{TeIL}{Tensor Intermediate Language}
\newacronym{tsv}{TSV}{through-silicon via}

\makeatletter
\lstdefinelanguage{mlir}{
    classoffset=0,
    morekeywords={
        module,
        func
    },
    morestring=[b]",
    alsoletter={\%},
    keywordsprefix={\%}
}
\makeatother

\begin{document}

\title{Automatic Creation of High-Bandwidth Memory Architectures from Domain-Specific Languages: The Case of Computational Fluid Dynamics}

\author{Stephanie Soldavini}
\affiliation{%
  \institution{Politecnico di Milano}
  \city{Milan}
  \country{Italy}
}
\email{stephanie.soldavini@polimi.it}

\author{Karl F. A. Friebel}
\affiliation{%
  \institution{Technische Universität Dresden}
  \city{Dresden}
  \country{Germany}
}
\email{karl.friebel@tu-dresden.de}

\author{Mattia Tibaldi}
\affiliation{%
  \institution{Politecnico di Milano}
  \city{Milan}
  \country{Italy}
}
\email{mattia.tibaldi@polimi.it}

\author{Gerald Hempel}
\affiliation{%
  \institution{Technische Universität Dresden}
  \city{Dresden}
  \country{Germany}
}
\email{gerald.hempel@tu-dresden.de}

\author{Jeronimo Castrillon}
\affiliation{%
  \institution{Technische Universität Dresden}
  \city{Dresden}
  \country{Germany}
}
\email{jeronimo.castrillon@tu-dresden.de}

\author{Christian Pilato}
\affiliation{%
  \institution{Politecnico di Milano}
  \city{Milan}
  \country{Italy}
}
\email{christian.pilato@polimi.it}

\renewcommand{\shortauthors}{Soldavini et al.}

\begin{abstract}
Numerical simulations can help solve complex problems. Most of these algorithms are massively parallel and thus good candidates for FPGA acceleration thanks to spatial parallelism. Modern FPGA devices can leverage high-bandwidth memory technologies, but when applications are memory-bound  designers must craft advanced communication and memory architectures for efficient data movement and on-chip storage. This development process requires hardware design skills that are uncommon in domain-specific experts.
In this paper, we propose an automated tool flow from a \gls{dsl} for tensor expressions to generate massively-parallel accelerators on HBM-equipped FPGAs. Designers can use this flow to integrate and evaluate various compiler or hardware optimizations. We use \gls{cfd} as a paradigmatic example.
Our flow starts from the high-level specification of tensor operations and combines an MLIR-based compiler with an in-house hardware generation flow to generate systems with parallel accelerators and a specialized memory architecture that moves data efficiently, aiming at fully exploiting the available CPU-FPGA bandwidth.
We simulated applications with millions of elements, achieving up to 103 GFLOPS with one compute unit and custom precision when targeting a Xilinx Alveo U280. Our FPGA implementation is up to 25$\times$ more energy efficient than expert-crafted Intel CPU implementations. 
\end{abstract}

\maketitle

\section{Introduction}

\textbf{Numerical simulations} are computationally-intensive applications that are used to solve many complex problems in industry~\cite{Schlatter2011}. These data-intensive applications are massively parallel since they use a combination of tensor operators to compute smaller and independent contributions that operate on different data to compose the final result~\cite{Swirydowicz2017}. In this context, \textbf{\gls{hpc}} is a powerful tool to reduce the costs of testing while giving the possibility of exploring more solutions. \gls{hpc} solutions can provide high-resolution physics results for many domains, including molecular dynamics~\cite{Chiu2010} or weather simulations~\cite{Sipkova2016}.

In this context, FPGA devices are increasingly used to achieve energy-efficient high performance by exploiting spatial parallelism with specialized accelerators. They are thus good candidates for accelerating numerical simulations.
Embedded FPGA devices can provide parallel architectures that overcome embedded processors but are limited in terms of resources and memory bandwidth~\cite{Friebel2021}. So, they cannot compete with large \gls{hpc} data centers. Instead, modern FPGA data center cards offer advanced \textbf{\gls{hbm} architectures} with multiple high-speed memory channels that enable efficient and parallel data transfers between the CPU and the reconfigurable logic~\cite{Weerasinghe2015,Kara2020}. \gls{hbm} is a modern memory technology that can offer a bandwidth of hundreds of Gigabytes per second~\cite{Jun2017}. For example, both Intel and Xilinx offer \gls{hbm} solutions in their FPGA devices: Intel Stratix 10 MX FPGAs include \gls{hbm}2 with 16GB of data that can be accessed up to 409~GB/s, while Xilinx Alveo FPGA cards offer 8GB of \gls{hbm}2 at 460~GB/s bandwidth. Such platforms enable the acceleration of memory-bound applications. Additionally, designers can use \textbf{\gls{hls}} tools to raise the abstraction level of hardware design~\cite{Nane2016,Lai2021}. However, designing efficient architectures for such systems is complex as it requires a concurrent optimization of communication, computation, and storage~\cite{Pilato2021}. These optimizations may be limited by platform constraints, like the physical architecture, which can make the routing stages more difficult, or the number of physical resources, which can limit the number of parallel executions
(see \autoref{sec:challenges} for further details on these kinds of challenges).

\begin{figure}
    \centering
    \includegraphics[width=0.7\textwidth]{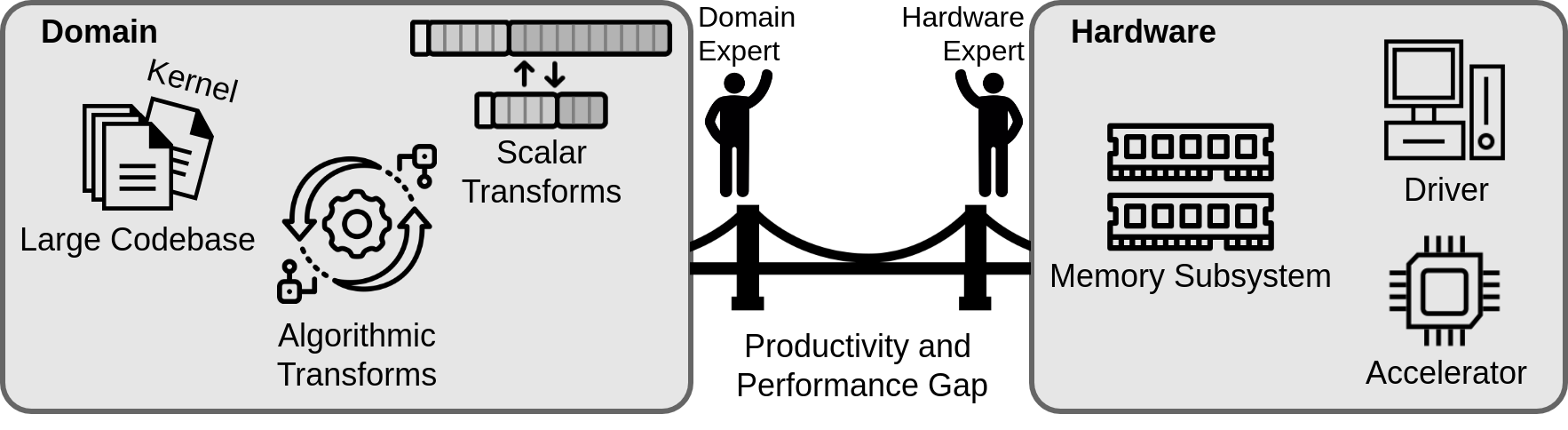}
    \vspace{-8pt}\caption{Our flow can help bridge the gap between domain and FPGA experts by providing a framework to match software and hardware requirements.}
    \label{fig:cartoon}
\end{figure}

Application and hardware developers face orthogonal challenges in fully exploiting \gls{hbm} architectures~\cite{Choi2020}. In \autoref{fig:cartoon}, we highlight some key problems addressed in this work on both sides, which create a gap between the experts in terms of:
\begin{itemize}[leftmargin=1em]
\item \textbf{productivity}: application designers usually have limited knowledge on hardware design and cannot create efficient hardware architectures to fully exploit the available FPGA technology. For this reason, they prefer to use \gls{dsl} descriptions to abstract the semantics of their operators;
\item \textbf{performance}: coordinating data transfers and execution requires an intimate knowledge of both the application and the target platform that is uncommon in many developers In this case, fine-tuned hardware descriptions are required to overcome these challenges.
\end{itemize}

\noindent On the architecture side, \gls{hbm} poses several challenges.
They have many parallel memory channels, but the overall number is limited (up to 32 in modern Xilinx Alveo cards). The communication cost between host and device memories is expensive: large CPU-FPGA data transfers can be much longer than the computation time of the kernel offloaded to the FPGA fabric. 

In this paper we look into computational kernels that are composed of tensor operations, using \gls{cfd} as a paradigmatic example.
Such tensor expressions are common across domains, including fluid dynamics, quantum chemistry, deep learning, image processing, and data analytics~\cite{Rink2018}, and put high pressure on FPGA resources (especially DSP and BRAM).
These resources are finite, and they can limit the deployment of parallel kernels. The designer must carefully trade-off kernel optimizations to achieve the best target architecture. 
In addition, the number of possible implementations of tensor expressions (see~\cite{brauckmann_pact21} for distribution of the speedup for simple loop programs) only exacerbates this problem.
Finally, accelerators demand efficient data movement to exchange data with the off-chip memory, but the designer needs to exploit the available bandwidth. Optimizing \gls{hls} code to maximize the performance is thus a complex task that requires modeling and exploring different solutions at both software and hardware levels~\cite{Siracusa2020}.

To address these challenges and support the designer in the optimization process, we propose a \textbf{DSL-to-bitstream workflow} composed of: (1) an \textbf{MLIR-based \gls{dsl} compiler} that abstracts low-level specification details (for enhancing designer's productivity by identifying the most suitable HLS-ready code to exploit hardware parallelism) and generates \gls{hls}-optimized codes for the kernels that enable us to deploy multiple hardware modules and to transfer the data more efficiently; and (2) an \textbf{automated \gls{hls}-based flow} that exploits this information to design architectures that can leverage the \gls{hbm} subsystem for efficient data transfers (for enhancing performance results). \textbf{Our workflow allows application designers to describe their applications in a high-level, domain-specific, and platform-agnostic language and use an automated toolchain to create the corresponding system architecture that exploits the intrinsic parallelism and the characteristics of \gls{hbm} systems.} The designer can select and apply various optimizations, enabling a non-FPGA expert to evaluate several alternatives.
This paper extends the work presented in~\cite{Friebel2021} to target FPGA data center cards with \gls{hbm} architectures.
Our novel flow is based on \gls{mlir}~\cite{Lattner2021} to progressively lower the specification without losing semantic information.
The rest of the flow creates the memory architecture (in C++) around the accelerator kernels, leveraging commercial \gls{hls} to generate the hardware description and, in turn, the FPGA bitstreams.
Our main contributions are:
\begin{itemize}[leftmargin=1em]
	\item we present a \textbf{compiler infrastructure based on \gls{mlir}} for a \gls{dsl} for tensor operations to automatically generate \gls{hls}-ready code for the computational kernels;
	\item we describe an \textbf{\gls{hbm}-oriented hardware generation flow} that interfaces with commercial \gls{hls} to generate an optimized system architecture; 
	\item we show how our flow can help an application designer in \textbf{optimizing the \textit{Inverse Helmholtz} operator}, a key element of \gls{cfd} simulations.
	This complex operator subsumes other widely-used tensor operators, like tensor contraction and tensor-based interpolation.
\end{itemize}
With our flow, the designer can generate, evaluate, and compare several alternatives, for example, trading off the accuracy of the results and performance.
Our analysis also provides useful guidelines for designers to understand what to do (or not to do) when implementing similar tensor-based applications on \gls{hbm}-based systems, especially when aiming at scaling up the computation with multiple computing units that execute in parallel.

\section{Background}\label{sec:background}

In this section, we present the main concepts at the basis of our work. We first introduce our target application (\autoref{sec:cfd}), highlighting that it is representative of other similar numerical applications. We then describe the characteristics and the challenges of our target platform (\autoref{sec:alveo} and \autoref{sec:challenges}, respectively). We conclude with an analysis of the related work (\autoref{sec:related}).

\subsection{CFDlang DSL for Spectral Element Methods}\label{sec:cfd}

% The inverse Helmholtz operator kernel code.
\begin{figure}[t]
\begin{minipage}[t]{0.55\textwidth}
\vspace{0pt}
\begin{lstlisting}[language=CFDlang,numbers=left,numbersep=5pt,basicstyle=\scriptsize\ttfamily,frame=tb,numberstyle=\tiny,linewidth=0.95\columnwidth,xleftmargin=2em,framexleftmargin=1.5em]
var input S   : [11 11]
var input D   : [11 11 11]
var input u   : [11 11 11]
var output v  : [11 11 11]
var t         : [11 11 11]
var r         : [11 11 11]
t = S#S#S#u . [[1 6][3 7][5 8]]
r = D * t
v = S#S#S#r . [[0 6][2 7][4 8]]
\end{lstlisting}
\vspace{-10pt}%
\captionof{figure}{DSL code for the Inverse Helmholtz operator ($p=11$).}%
\label{fig:helm_op:dsl}%
\end{minipage}
\vspace{-6pt}
\end{figure}

In numerical mathematics, \gls{sem} are common in solving partial differential equations (PDEs), like the Navier-Stokes equations~\cite{Maday1989}, which cannot be solved analytically. \gls{sem} approximates the solution by transforming the unknown physical quantities of the problem into spectral coefficients.
To reduce the numerical complexity, the simulated volume is divided into~$N_{eq}$ smaller \textbf{elements}.
To further reduce the error, \gls{sem} uses an approximation based on polynomials of a higher degree ($p>1$).
The solution is expressed as a linear system of equations that can be solved locally for each element.
An element solution~$e$ can be represented in three dimensions as a tensor~$v_{ijk, e}$ with~$i,j,k \in \{0, \dots, p\}$. Often, the polynomial degree~$p$ is the same for all spatial dimensions. All elements operate on independent tensors and can be elaborated in parallel.

In our example, we focus on solving the Helmholtz equation $\lambda u - \nabla^2 u = f$ for quadrilateral elements. The Inverse Helmholtz operator subsumes simpler operators (e.g., interpolation) that are similarly relevant in \gls{cfd} simulations~\cite{Huismann2017b}.
The operator can be formulated as:
\begin{subequations}
  \begin{align}
    \label{eq:helm_op:1}
    t_{ijk,e} &= \sum_{l=0}^{p} \sum_{m=0}^{p} \sum_{n=0}^{p} S^T_{li} \cdot S^T_{mj} \cdot S^T_{nk} \cdot u_{lmn,e} = {\left(S \otimes S \otimes S \otimes u_e\right)_{iljmkn}}^{lmn} %\nonumber 
    \\
    \label{eq:helm_op:2}
    r_{ijk,e} &= D_{ijk,e} \cdot t_{ijk,e}\\
    \label{eq:helm_op:3}
    v_{ijk,e} &= \sum_{l=0}^{p} \sum_{m=0}^{p} \sum_{n=0}^{p} S_{li} \cdot S_{mj} \cdot S_{nk} \cdot r_{lmn,e} = {\left(S \otimes S \otimes S \otimes r_e\right)_{limjnk}}^{lmn} %\nonumber
  \end{align}
\end{subequations}

The CFDlang \gls{dsl} for tensor operations~\cite{Rink2018} enables us to encode these kinds of operators concisely.
We have chosen this particular DSL because of its excellent domain capture and limited optimization scope.
CFDlang is target-agnostic and its user interface is close to the mathematical problem specification, reducing the programming burden on application developers. \autoref{fig:helm_op:dsl} shows a CFDlang program implementing the Inverse Helmholtz operator, where lines 7-9 are the direct transcriptions of the expressions \eqref{eq:helm_op:1}, \eqref{eq:helm_op:2} and \eqref{eq:helm_op:3}.
The CFDlang description does not define exact implementation details, allowing the compiler to optimize the operations for a given target.
Another implicit part of the given DSL program is that it is assumed to be applied to all the independent elements in an implicit outer \textquote{element loop}.

From the application developer viewpoint, once the kernel is specified in CFDlang, it can be integrated into larger Fortran or C++ code applications via their respective \gls{ffi} mechanisms where we can also embed the FPGA runtime library calls.

Previous work found that typical software implementations achieve performances between 1~and~16 GFLOPS based on the polynomial degree $p$, with an average power consumption of at least 100W~\cite{Rink2018}. GPUs, in turn, do not feature as good a scaling behavior for these kernels~\cite{Huismann2017a}.

\subsection{HBM-based Platform Description}\label{sec:alveo}

\begin{figure}[t]
\begin{minipage}[t]{0.48\textwidth}
\vspace{0pt}
\centering
\includegraphics[width=0.85\textwidth]{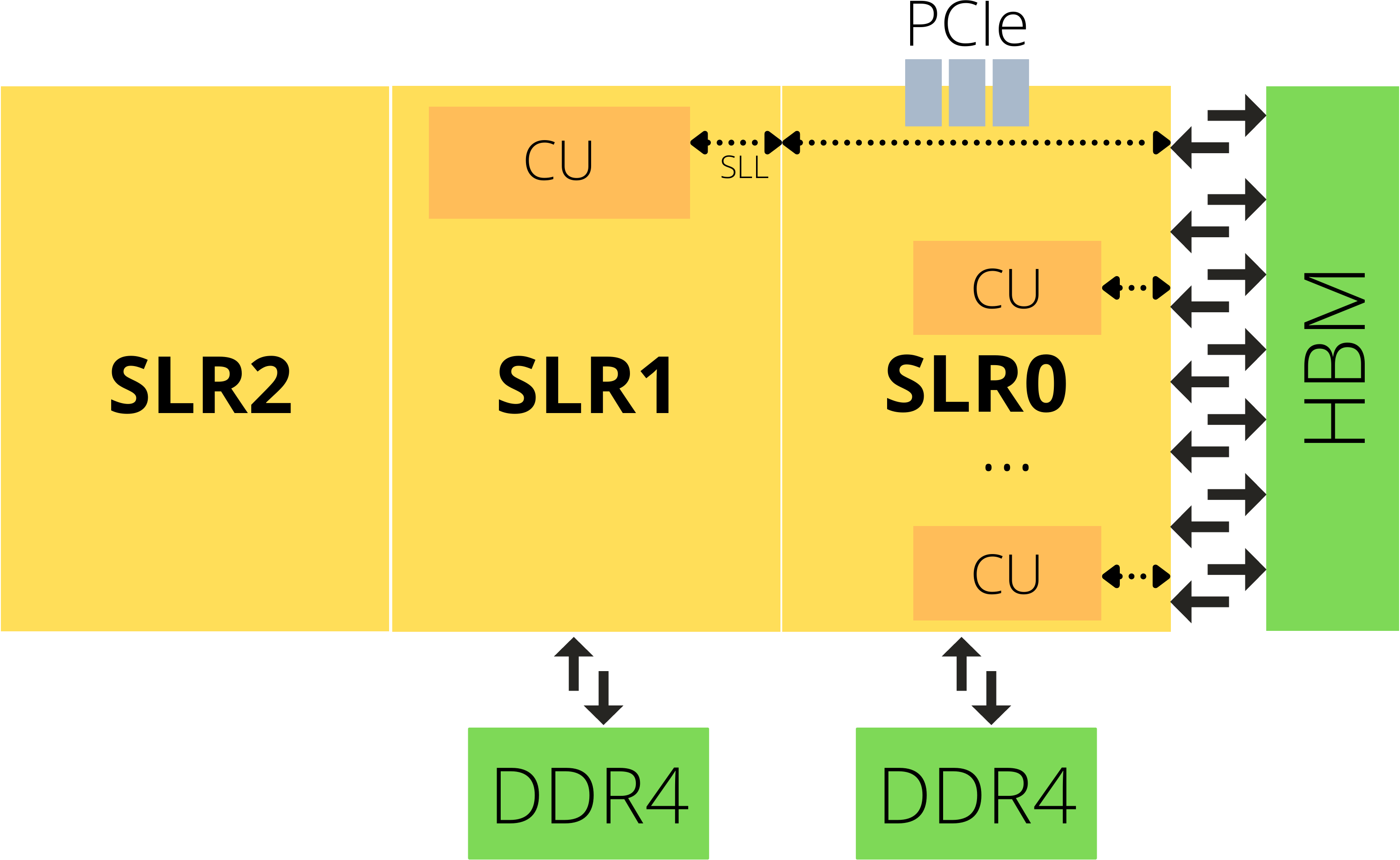}
\captionof{figure}{Architecture of the XCU280 device.}\label{fig:floorplan}
\end{minipage}\hspace{4pt}
\begin{minipage}[t]{0.46\textwidth}
\centering
\footnotesize
\vspace{0pt}
\captionsetup{type=table}
\captionof{table}{Alveo U280 SLR resources.}\label{tab:slr_resources}
\vspace{-10pt}\begin{tabular}{@{}c@{}ccc@{}}\toprule
 \textbf{Resources} & \textbf{SLR0}  & \textbf{SLR1} & \textbf{SLR2}\\\midrule
 HBM & 32$\times$256MB & - & - \\
 DDR4 & 16GB & 16GB & - \\
 PLRAM & 2$\times$4MB & 2$\times$4MB & 2$\times$4MB \\\midrule
 CLB LUT & 369K & 333K & 367K \\
 CLB Register & 746K & 675K & 729K \\
 Block RAM tile & 507 & 468 & 512 \\
 UltraRAM & 320 & 320 & 320 \\
 DSP & 2,733 & 2,877 & 2,880 \\\bottomrule
\end{tabular}
\end{minipage}
\vspace{-9pt}
\end{figure}

\gls{hbm} is a novel memory architecture that enables high-performance and adaptability for memory-bound applications~\cite{Fujita2021}. \gls{hbm} is a \textbf{3D-stacked DRAM} that offers high-bandwidth and energy-efficient data movements. The logic die is connected to the HBM die(s) with \glspl{tsv}. The rest of this work focuses on the \textbf{Xilinx Alveo U280} data center accelerator card. 
However, other \gls{hbm}-based FPGA devices (like the Intel Stratix 10 MX) are conceptually similar~\cite{Choi2020}.
The Alveo U280 is built on the Xilinx 16nm UltraScale+ architecture and offers a rich set of memory solutions, as shown in \autoref{fig:floorplan}.
The Alveo U280 card features the XCU280 FPGA, which combines three \glspl{slr}. An \gls{slr} is a physical section of the FPGA with a specific amount of resources and connections, as shown in \autoref{tab:slr_resources}. SLR0 integrates an \gls{hbm} controller to interface with the \gls{hbm}2 subsystem through 32 \textbf{\glspl{pc}} each with direct access to 256~MB of storage (8~GB in total). Each 256-bit \gls{pc} operates at 450~MHz, yielding a maximum bandwidth of 14.4~GB/s. The full system can thus achieve a theoretical bandwidth of 460.8~GB/s. 
SLR0 also connects to the host via 16 lanes of the PCI Express (PCI-e) interface. SLR0 and SLR1 each connect to 16~GB of DDR4 each. Finally, each region has up to 8~MB of PLRAM for fast access to small data sets.
In the following, we will use the term \textbf{global memory} for the set of memories available on the board. The host must transfer data into the device global memory before they can be accessed by the FPGA logic.

The target system for the Alveo U280 is composed of multiple \textbf{\glspl{cu}}.
Each \gls{cu} is a user-defined hardware module that can be attached to any of the \glspl{pc} through independent AXI interfaces, while the built-in \gls{hbm} controller and switch have access to all physical channels. The \gls{cu} can be described in C++ and synthesized with \gls{hls} or specified directly in RTL. Multiple \glspl{cu} allow parallel execution but must be connected to different \gls{hbm} channels. The \textbf{system configuration file} describes the connections between the \gls{cu} ports and the \gls{hbm} channels. The required logic is automatically generated during system synthesis.

Xilinx offers a \textbf{unified software platform}, called Vitis, to develop FPGA applications. Vitis includes a rich set of hardware-accelerated open-source libraries optimized for Xilinx FPGA and the Xilinx Runtime library (XRT) to facilitate communication between the host application (running on the host CPU) and the accelerator deployed on the reconfigurable portion of the card, which is connected via PCI-Express. It also includes user-space libraries and APIs, kernel drivers, and board utilities that can be used to measure performance and monitor power consumption. In this work, we aim to automate the generation of \gls{cu} descriptions and the associated configuration file directly on top of the existing Xilinx libraries.

\subsection{Challenges for Efficient HBM-Optimized Designs}\label{sec:challenges}

In modern FPGA data center cards like the Xilinx Alveo U280, DDR4 memory is excellent for accessing large data sets with modest latency, but the transfer bandwidth is limited to 36 GB/s and no more than two parallel accesses~\cite{Huang2021}. Conversely, \gls{hbm} has slightly higher latency, but it can reach a much higher theoretical bandwidth (460.8 GB/s) when fully using all its 32 \glspl{pc}~\cite{Choi2020,Singh2021}.
\gls{hbm}-based architectures are becoming popular for memory-bound applications~\cite{Kara2020,Calore2021}. They have been used for convolutional neural networks, graph analytics, or weather-prediction~\cite{Venkataramanaiah2020, Liu2021, Sgherzi2021, Singh2020}. These platforms also feature PLRAM which can be used for small amounts of frequently-accessed data. In the case of the Alveo U280, the designer can configure these connections in the system configuration file. However, several factors can affect the performance of the resulting system.

\noindent\textbf{Challenge 1: CPU-Host Communication Cost. }
Moving data across the PCI-e interface to the global memory has a high cost compared to the data access from the logic regions to the global memory. So, such movements must be optimized to reduce their latency. First, the designers must move enough data to perform significant computation, for example by operating on multiple elements in sequence before sending back the results.
In many cases, the kernel must read the same data multiple times. For example, the Inverse Helmholtz operator must read matrix \textit{S} several times during the two tensor contractions. To avoid multiple global memory accesses to the same data, designers can use internal buffers that gather enough data. 

\noindent\textbf{Challenge 2: Read/Write Burst Transactions. }
Vitis can automatically infer the size of data transfers to create more efficient burst transactions. However, frequently switching between read and write transactions is inefficient due to memory controller timing parameters. So, transactions should be ordered to maximize the data movement in one direction before switching to moving data in the other direction. The alternative is to partition reads and writes into separate \gls{hbm} channels, at the cost of increasing the number of channels required for each \gls{cu}.

\noindent\textbf{Challenge 3: Full Bandwidth Utilization. }\label{sec:bandwidth_utilization} The AXI interfaces to the \gls{hbm} channels can be configured with wider data busses. For example, the designer can use a bus width of 256 bits running at 450~MHz or even 512~bits running at 225~MHz. This allows the \gls{cu} to exchange more data in each clock cycle, reducing the read/write cost from/to the \gls{hbm} channels. For example, when configuring the 32 parallel \gls{hbm} channels with a bus width of 256 bits, the potential bandwidth between the CPU and the FPGA is about 460~GB/s at 450~MHz. However, such data must be processed efficiently and in parallel to avoid performance bottlenecks~\cite{Kara2020}.

\noindent\textbf{Challenge 4: HBM-Data Allocation. } To avoid congestion in the switch, it is important to partition the data into the different \gls{hbm} memory regions such that each \gls{cu} uses as few channels as possible and shares these channels with as few other \glspl{cu} as possible. 
Otherwise, the designer must introduce an efficient crossbar to mitigate congestion~\cite{Choi2021}.

\noindent\textbf{Challenge 5: Synthesis-Related Issues. }
FPGA devices include hard macros like DSP and BRAM for efficient computation and storage, respectively. However, the physical location of these FPGA resources can affect routing. Also, the XCU280 FPGA, which is the device at the basis of the Alveo U280, is partitioned into three \glspl{slr}, but only SLR0 is connected to the \gls{hbm} channels (cf. \autoref{sec:alveo}). When the \glspl{cu} cannot fit into a single \gls{slr}, they are automatically split over multiple \glspl{slr} using special resources called \gls{sll} routes. These resources allow routing between \glspl{cu} in any \gls{slr}, enabling access to all memory resources, but they introduce performance overhead.
So, if a \gls{cu} requires access to the \gls{hbm}, it should be allocated in SLR0. Also, hardware modules that access multiple \gls{hbm} channels should be close together to reduce wire length.

\subsection{MLIR}\label{sec:MLIR}

\gls{mlir}~\cite{Lattner2020} is a recent compiler infrastructure that allows defining custom abstractions and high-level transformations.
It has received considerable traction as a framework for developing domain-specific compilers for heterogeneous systems, especially for machine learning frameworks.
\gls{mlir} is in itself not a fixed \gls{ir}, but an infrastructure with a central plug-in mechanism to extend the vocabulary of the \gls{ir} using \emph{dialects}.
A dialect implements an intermediate abstraction as a set of operations, types, and attributes.
Examples of prominent dialects include \texttt{linalg} for linear algebra operations, \texttt{affine} for nested loop programs, and the \emph{\texttt{llvm}} dialect which enables a seamless transition into the LLVM compiler framework at the lowest level of abstraction. 
Custom dialects allow compiler designers to develop analyses and transformations at the right level of abstraction.
Dialects can be integrated into larger language stacks, fostering the reuse of abstractions and code transformations. 
A typical way of integrating a dialect is using \emph{lowering}, where a more abstract dialect is converted into a more concrete one through specific transformations.
Arguably the most complete flow built on top of \gls{mlir} to date is presented in~\cite{Vasilache2022}, where authors describe how \gls{mlir} allows for a decentralized and composable compiler. 

In this paper we leverage \gls{mlir} to create a composable and reusable domain-specific compiler. 
We implement the abstractions for tensors described in \autoref{sec:cfd} and propose a lowering pipeline that integrates with existing dialects in \gls{mlir}, while exposing profitable transformations for HBM-optimized designs.
We also show how hardware-specific abstractions, e.g., for number representations, are integrated into the compilation flow.

\subsection{Related Work}\label{sec:related}

As FPGA devices are entering data centers, \gls{hbm} architectures are becoming extremely popular to accelerate a variety of data-intensive applications~\cite{Jun2017}. Benchmarking these architectures defined that HBM provides higher bandwidth and access latency than traditional DDR~\cite{Venkataramanaiah2020, Huang2021, Meyer2020, Lu2020}. Instead, comparing FPGA performance with both CPUs and GPUs in HPC workloads concluded that there is potential in this technology. They also point out that although FPGAs struggle to compete in absolute terms with GPUs, they achieve much greater energy efficiency~\cite{Calore2021,Nguyen2020,Kuramochi2020}. 
%%%
\cite{Fujita2021} and \cite{Choi2021} argue that one of the main problems in FPGA is the absence of a comprehensive memory hierarchy, like the one in CPUs. For this reason, they propose two different custom crossbar solutions that efficiently access the HBM and create an additional memory layer using the BRAMs available inside the FPGA. However, these solutions restrict the amount of BRAMs available to implement the user's application and they do not fully exploit the FPGA pipeline parallelism. Other works like \cite{Holzinger2021,Choi2020,Singh2020} define a series of guidelines to achieve high performance. Specifically, the designer must always instantiate the maximum number of computing units to exploit HBM parallelism, use 256-bit packets to maximize the bandwidth of the HBM AXI port, exploit the algorithm parallelism through hardware pipelining, and finally execute the task in a dataflow manner \cite{Kara2020}. However, none of these works evaluated these optimizations together, along with the effects when scaling the number of computing units.
None of the papers presented so far defines methodologies at the HLS level that can be applied to design to improve its performance, especially for tensor applications.
\cite{Cong2017} proposes three buffer restructuring approaches, coarse-grained, fine-grained, and hybrid, to adjust the BRAM and LUT utilization as needed. Similarly, tensor optimizations were proposed in~\cite{Siracusa2020}.
\cite{Rajagopala19} studies the effect of HLS pragma directives on the design. In particular, they indicate that loop unrolling, and array partitioning directives can cause overhead in off-chip memory performance due to non-burst access. Finally, they provide a code transformation technique to exploit the data width of the memory controller. The Merlin compiler\footnote{\url{https://github.com/Xilinx/merlin-compiler}} provides an infrastructure for source-to-source transformation to accelerate applications on FPGA that are already described in C++~\cite{CongHPW016}. Instead, operating at the \gls{dsl} level allows us to implement high-level transformations to the source code, including exploring different memory layouts.

% Tensor DSLs and their compilers.
The core of our \gls{dsl} abstraction consists in modeling tensor expressions.
There are several such \glspl{dsl}, like \gls{tc}~\cite{Vasilache2018} and TensorFlow~Eager~\cite{Agrawal2019}, that are capable of representing tensor operations in a more general domain.
However, many are based on backing software libraries, such as TensorFlow~\cite{Abadi2016} and Theano~\cite{Bergstra2010}, which are less flexible in terms of their back end.
TVM~\cite{Chen2018} is a compiler focusing on \gls{ml} that is not as limited in that regard, even providing an FPGA back-end.
Its focus on \gls{ml} makes it hard to apply TVM to other application domains. 
A similar application-specific flow is provided by ESP4ML~\cite{Giri2020}, which explores a bigger design space of systems-on-chip (SoCs) using \gls{hls} but does not provide automatic generation from tensor expressions.
More generally applicable, the Halide~\cite{Ragan-Kelley2013} ecosystem provides extensive support for heterogeneous systems. Spatial~\cite{spatial} provides automatic compilation and system generation from high-level descriptions. The framework offers high-level communication libraries, but the application development burden is still on the designer.

% MLIR.
Thanks to \gls{mlir}~\cite{Lattner2021}, as a framework for designing intermediate languages, several projects have recently proposed programming flows for heterogeneous systems. 
In the \gls{mlir} infrastructure, compiler internals are composed using a plug-in system, which allows connecting various front-, middle- and back-ends, which may be entirely orthogonal.
For example, with Teckyl\footnote{\url{https://github.com/andidr/teckyl}}, the aforementioned \gls{tc} has received a front-end for the \gls{mlir} compiler infrastructure, enabling a variety of back-end consumers.
As an alternative to just lowering to LLVM~\cite{Lattner2004},
IREE\footnote{\url{https://google.github.io/iree/}} offers perhaps the first functional end-to-end flow entirely designed within \gls{mlir}.
% Pluggable MLIR stuff.
Closer to our domain is the recent Open Earth Compiler~\cite{Gysi2021} for efficient GPU code generation for stencil operations.
We also rely on the plug-and-compile \gls{mlir} philosophy, which will allow others to reuse our abstractions and allows us to profit from existing language stacks and compilation flows.

% Polyhedral stuff.
In the context of \gls{mlir}, projects such as Polygeist~\cite{Moses2021} and Phism~\cite{Zhao2021} use polyhedral modelling to ingest, optimize, and even emit C code for \gls{hls}.
Our compiler still employs polyhedral analysis and rescheduling as described in~\cite{Friebel2021}.
Streaming implementations are a challenge in FPGA implementations, which can tie into the polyhedral model via consecutivity constraints~\cite{Verdoolaege2017}.

In conclusion, to address productivity issues in scientific computing, designers need a comprehensive framework that 1) allows them to use high-level tensor expressions to specify the behavior, 2) automates the hardware generation process on top of commercial \gls{hls} solutions, and 3) efficiently targets modern \gls{hbm} architectures.

\section{Automatic CFDlang-to-Bitstream Flow}\label{sec:dslflow}

\begin{figure}
\includegraphics[height=5cm]{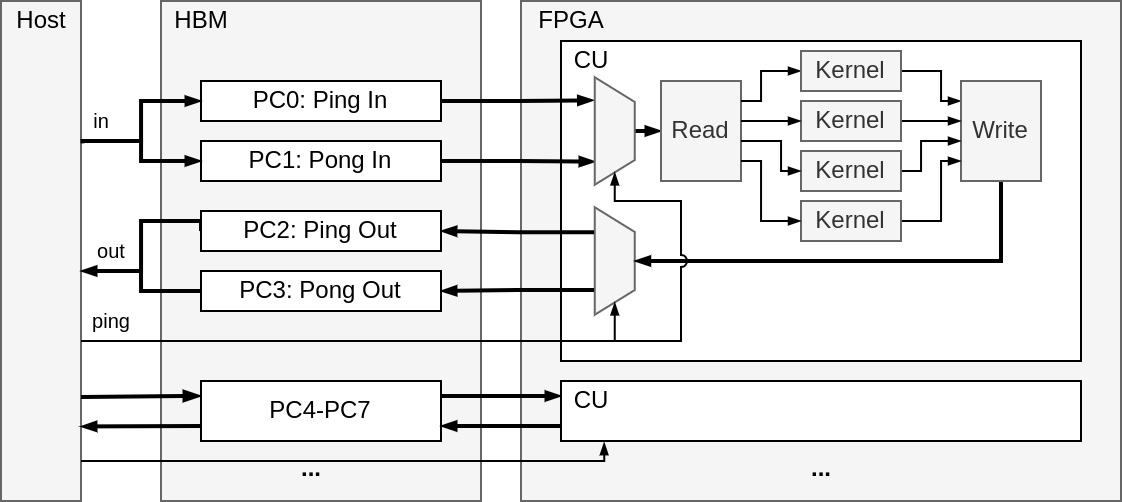}
\caption{Target system for the massively-parallel applications, like \glsentryfull{cfd}.}\label{fig:target_system}
\vspace{-6pt}
\end{figure}

This section describes our proposed approach. We first discuss our envisioned target architecture to implement \gls{cfd} applications on \gls{hbm} architectures (\autoref{sec:system}). We then outline our CFDlang-to-Bitstream flow (\autoref{sec:flow_overview}), followed by details on the two major components: the CFDlang compiler (\autoref{sec:compiler_arch} and \autoref{sec:compiler}) and the hardware generation flow (\autoref{sec:mem} and \autoref{sec:mem_opt}).

\subsection{Target System Architecture}\label{sec:system}

To simulate the complete volume, the \gls{cfd} application applies the \textit{Inverse Helmholtz} operator to $N_{eq}$ independent elements. To exploit such intrinsic parallelism and address the challenges discussed in \autoref{sec:challenges}, we aim to build a target system like the one in~\autoref{fig:target_system}.

From the system-level perspective, the \gls{cfd} simulation runs on the host CPU, which sends the data to the FPGA HBM via PCIe. Once the data are in the HBM, the \glsentryfull{cu} can fetch them in parallel through AXI channels for several simulation elements. Since each \gls{hbm} interface is 256 bits wide, the channel can be conceptually divided into multiple lanes based on the data bitwidth. Each lane can serve an independent accelerator, called \textit{Kernel}, derived from the high-level \gls{dsl} description, i.e., the Inverse Helmholtz operator of \autoref{fig:helm_op:dsl}. 
The \gls{cu} is the fundamental unit that can be directly attached to the \gls{hbm} channels and can be internally composed of several kernels with a wide range of communication patterns depending on the high-level description of the functionality to be implemented. 
To manage data exchanges between the HBM interfaces and the parallel kernels, the \gls{cu} uses two additional hardware modules, i.e., \textit{Read} and \textit{Write} in \autoref{fig:target_system}, that execute in parallel to the kernels and communicate with a dataflow model. The \textit{Read} module fetches the input data of one element (matrices \texttt{S}, \texttt{D}, and \texttt{u}) from the \gls{hbm} memory into internal buffers, while the \textit{Write} module transfers the corresponding results (matrix~\texttt{v}) into the \gls{hbm} memory. To avoid conflicts and reduce the complexity of the AXI interfaces, we assign these modules to independent \gls{hbm} channels.
The data read and write modules within the \gls{cu} split the 256-bit \gls{hbm} data into 32-bit or 64-bit data for the computation modules based on the data format.

To hide the communication cost, the host computes the maximum number of sets of input data that can fit in one \gls{hbm} channel. We thus define a \textbf{batch} as the number of elements $E$ that the \gls{cu} kernels will elaborate before the host retrieves the output. Executing a batch of several elements maximizes the size of the host data transfers, minimizing their associated overhead. In this way, the data exchanges between the CPU and the global memory can be overlapped with the \gls{cu} execution using a double-buffer method. While the host is exchanging data with the \textit{Ping} (\textit{Pong}) channel, \glspl{cu} can operate on the \textit{Pong} (\textit{Ping}) channel. Given the \gls{cu} structure, we can compute the number of batches $N_b=N_{eq}/E$ to be executed based on the total number of elements $N_{eq}$ to be simulated.

To further exploit parallelism, we can then replicate the \gls{cu} structure multiple times, each of them operating on independent \gls{hbm} channels. Let $N_{cu}$ be the number of \glspl{cu} that can be instantiated based on the resource constraints, the host application executes $I=N_b / N_{cu}$ iterations, evenly distributing the number of batches over the available \glspl{cu}.

In the following, we describe our DSL-to-bitstream flow to automatically create this architecture on top of the existing HLS platform (e.g., Vitis for the Alveo U280). We also describe how to automate the integration of the optimizations to increase performance and resource efficiency and how to modify the host code to match the memory layout required by the hardware modules.

\subsection{Overview of the Proposed Flow}\label{sec:flow_overview}

\begin{figure}
    \includegraphics[width=0.98\linewidth]{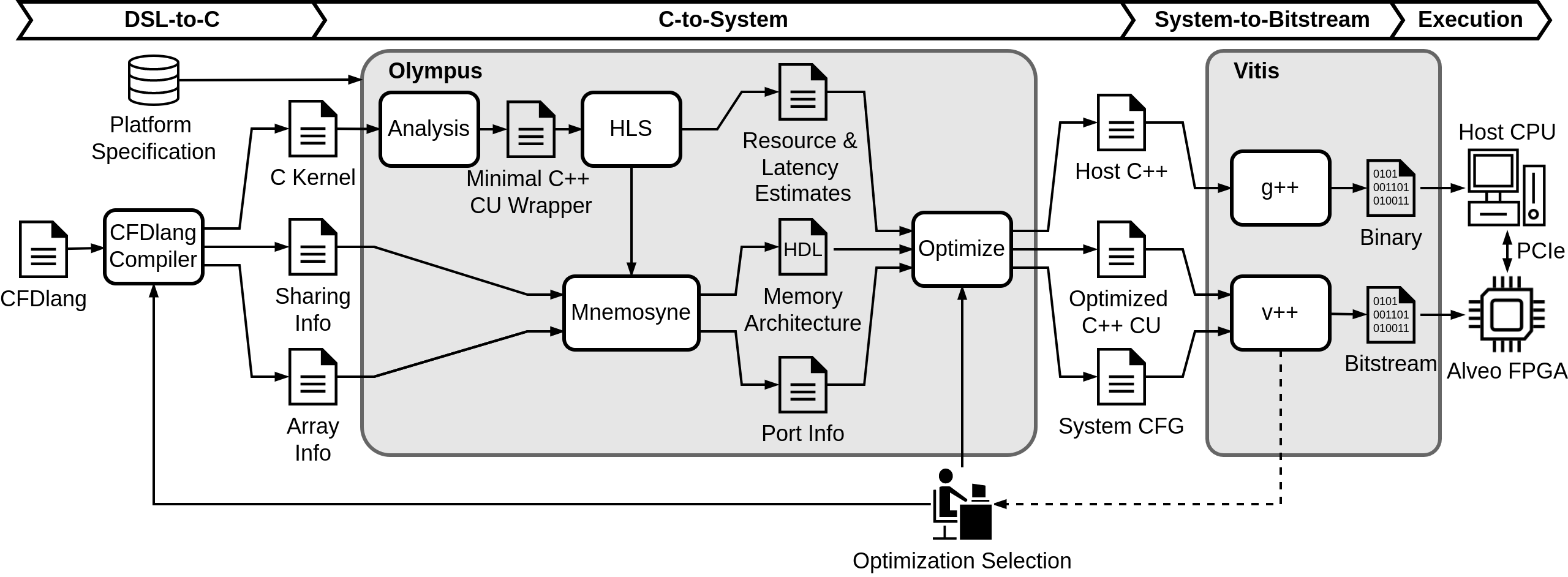}
    \vspace{-6pt}
    \caption{Tool flow from CFDlang to FPGA bitstream generation.  }\label{fig:toolflow}
\end{figure}

We propose a modular tool flow that simplifies the creation of FPGA accelerators for tensor-based applications, like numerical simulations, that are expressed in domain-specific languages.
\autoref{fig:toolflow} shows an overview of our flow which is composed of two major steps. We start from the \gls{dsl} description of the kernel and generate the corresponding optimized and HLS-friendly C code (\textbf{DSL-to-C generation}).
The compiler performs a series of transformations at different levels of abstraction (cf. \autoref{sec:compiler}), inserts pragmas to guide the \gls{hls} flow, and produces metadata for the memory generation.
Starting from the C code of the kernel, we create the corresponding parallel system with multiple \glspl{cu} (cf. \autoref{sec:mem}), each of them connected to one or more \glspl{pc} (\textbf{C-to-system generation}). Also, each \gls{cu} can instantiate one or more kernels based on the required amount of FPGA resources (cf. \autoref{sec:mem_opt}), along with the logic to move data from the global memory to the kernel on-chip buffer (\textit{Read} and \textit{Write} modules). We specify the \gls{cu} functionality in C++, wrapping the kernel code with specific code and HLS directives to enable the generation of the memory-optimized architecture.
To do so, we use a combination of distinct tools: a redesign of the \textbf{CFDlang} compiler (cf. \autoref{sec:compiler_arch}), which integrates novel kernel-level transformations, the \textbf{Mnemosyne} tool~\cite{Pilato2017}, which improves memory sharing \text{inside} the single kernel to reduce on-chip memory requirements, and the newly-developed \textbf{Olympus} tool, which generates the system architecture (i.e., \gls{cu} description and configuration file) around the kernels based on the optimization requested by the designer. Olympus also generates the corresponding host software to control the accelerators. This code comprises multiple steps (data allocation, kernel configuration, data transfers, synchronization primitives). Each step is wrapped into a specific function so that it can be further specialized in the hardware generation flow with specific optimizations.
We then use the \texttt{v++} tool of the commercial Xilinx Vitis Unified Platform to create the hardware description with HLS and generate the final FPGA bitstream. The Vitis platform also builds and links the host application with the Xilinx Runtime (\texttt{g++}). Our flow can easily be extended to target similar \gls{hbm}-based platforms (like the Intel Stratix 10 MX) with the respective toolchains. 

\subsection{Compiler Architecture}\label{sec:compiler_arch}

% Introduction.
Compiling for the template architecture in \autoref{fig:target_system} is fundamentally different to compiling for mainstream multicore CPUs.
Instead of extending the original CFDlang compiler~\cite{Rink2018} with new target support, we devise an entirely new implementation based on the \gls{mlir} framework.
With \gls{mlir}, we raise the language abstraction level while remaining target-agnostic in our front end.
At the same time, our middle-end is also mostly agnostic towards the concrete \gls{dsl} we chose.

\begin{figure}[tbp]
    \centering
    \includegraphics[width=0.8\columnwidth]{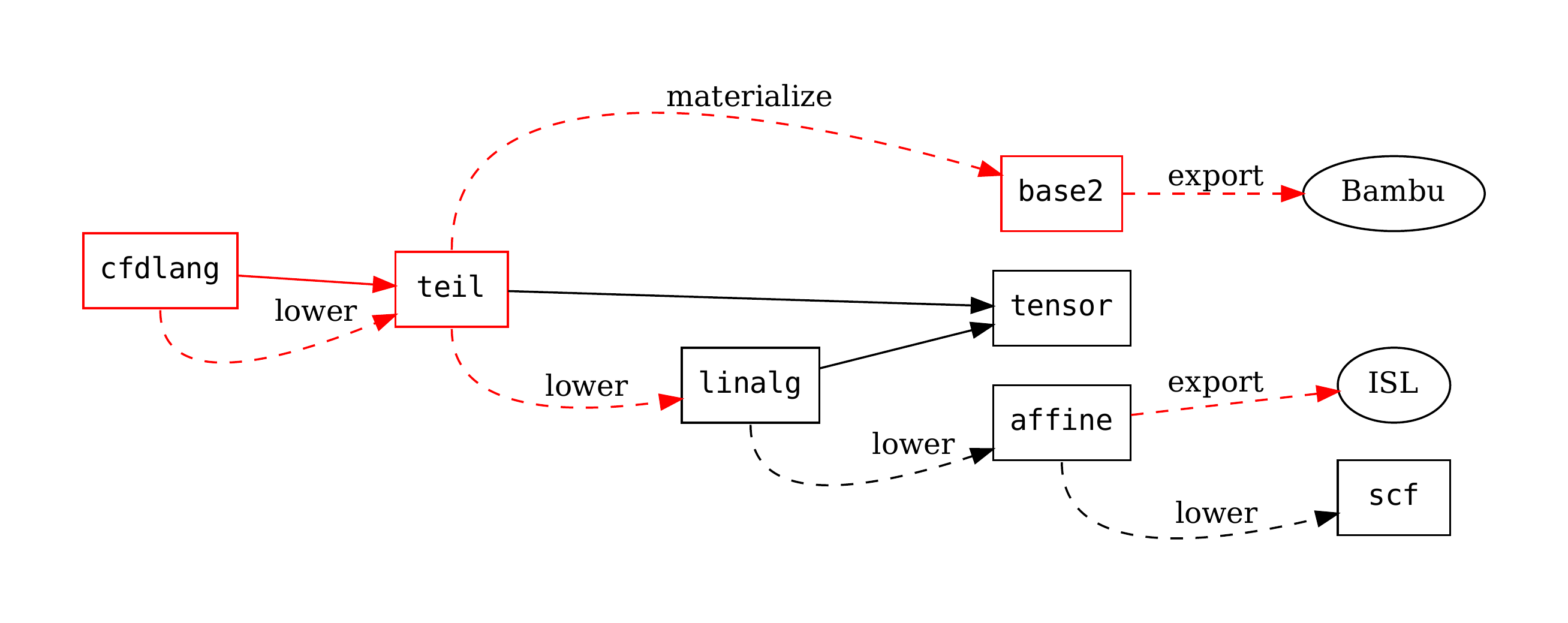}
    \vspace{-22pt}\caption{\gls{mlir} dialects and tools (ellipses) in the compiler and their dependencies. 
    Our contributions to the \gls{mlir} language stack are 
    marked in \textcolor{red}{red}.}\vspace{-6pt}\label{fig:dialectDeps}
\end{figure}

% MLIR framework.
This also allows us to profit from the well-engineered and widespread \gls{mlir} and LLVM ecosystems,
enabling re-use of abstractions and of lowering flows to different architectures.
In the following, we describe the new CFDlang compiler infrastructure, focusing on the stack of dialects we created to support \gls{cfd} simulations (cf. \autoref{fig:dialectDeps}).
Our infrastructure includes the key additions of only three new dialects: \texttt{cfdlang}, i.e., our language front-end, \texttt{teil}, i.e., our domain optimization middle-end, and \texttt{base2}, i.e., our example for a hardware-specific back-end task.
Apart from the dialects \texttt{linalg} and \texttt{affine} mentioned in \autoref{sec:MLIR}, the figure also includes other standard dialects, like \texttt{tensor} for cross-domain tensor operations, \texttt{scf} to model structured control flow with, e.g., explicit loops, and external tools such as the ISL, which we use to produce the code for the downstream \gls{hls} compilation.
The transformations performed within these dialects are discussed in \autoref{sec:compiler}.

Since we aim at leveraging a commercial \gls{hls} tool that does not currently support MLIR as input, our DSL compiler is an MLIR-based transpiler that emits C99 source code and additional metadata outputs to directly interface with Vitis HLS and Mnemosyne, respectively.

\subsubsection{The Front-end}

% Motivation.
Following our motivating example given by \autoref{fig:helm_op:dsl}, we enter the compiler flow at the highest level of abstraction.
Here, the \texttt{cfdlang} dialect replaces both the \gls{ast} and expression tree representations that were used previously in CFDlang~\cite{Rink2018}.
During translation, the \gls{dsl} parser directly emits \texttt{cfdlang} dialect operations while ingesting the \gls{dsl} code, resulting in \autoref{fig:dialect:translation}.
This dialect mirrors the syntactical elements of the \gls{dsl} in \gls{mlir}.
To establish a working front-end, we implement a translation from the \gls{dsl} to the \texttt{cfdlang} dialect, which also works backward.
In addition, we implement a lowering to our \gls{dsl} agnostic middle-end dialect \texttt{teil}.
We benefit from the \gls{mlir} diagnostic engine and immediate semantic analyses and verification in the \gls{mlir} infrastructure.

\begin{figure}
    \centering
    \begin{subfigure}{0.9\columnwidth}
        \centering
\begin{lstlisting}[language=mlir,basicstyle=\tiny\ttfamily,linerange={5-14},breaklines=true,postbreak=\mbox{\textcolor{red}{$\hookrightarrow$}\space},frame=tb,linewidth=\columnwidth]
    // t = S#S#S#u . [[1 6][3 7][5 8]]
    cfdlang.define @t : [11 11 11] {
        %S = cfdlang.eval @S : [11 11]
        %u = cfdlang.eval @u : [11 11 11]
        %0 = cfdlang.prod %S, %u : [11 11], [11 11 11]
        %1 = cfdlang.prod %S, %0 : [11 11], [11 11 11 11 11]
        %2 = cfdlang.prod %S, %1 : [11 11], [11 11 11 11 11 11 11]
        %3 = cfdlang.cont %2 : [11 11 11 11 11 11 11 11 11] indices [2 7][4 8][6 9]
        cfdlang.yield %3 : [11 11 11]
    }
\end{lstlisting}
        \vspace{-10pt}\caption{Translation of \autoref{fig:helm_op:dsl} (excerpt)\vspace{6pt}}\label{fig:dialect:translation}
    \end{subfigure}
    \begin{subfigure}{0.9\columnwidth}
        \centering
\begin{lstlisting}[language=mlir,
        numbers=left,numbersep=-5pt,
        basicstyle=\tiny\ttfamily,linerange={3-11},breaklines=true,postbreak=\mbox{\textcolor{red}{$\hookrightarrow$}\space},frame=tb,linewidth=\columnwidth]
        %0 = teil.prod %S, %u : tensor<11x11x!teil.num>, tensor<11x11x11x!teil.num>
        %1 = teil.diag 2 5 %0 : tensor<11x11x11x11x11x!teil.num>
        %2 = teil.red add 2 %1 : tensor<11x11x11x11x!teil.num>
        %3 = teil.prod %S, %2 : tensor<11x11x!teil.num>, tensor<11x11x11x!teil.num>
        %4 = teil.diag 2 5 %3 : tensor<11x11x11x11x11x!teil.num>
        %5 = teil.red add 2 %4 : tensor<11x11x11x11x!teil.num>
        %6 = teil.prod %S, %5 : tensor<11x11x!teil.num>, tensor<11x11x11x!teil.num>
        %7 = teil.diag 2 5 %6 : tensor<11x11x11x11x11x!teil.num>
        %8 = teil.red add 2 %7 : tensor<11x11x11x11x!teil.num>
\end{lstlisting}
        \vspace{-10pt}\caption{Optimized \texttt{teil} lowering of \autoref{fig:dialect:translation}}\label{fig:dialect:lowering}
    \end{subfigure}
    \vspace{-5pt}\caption{\gls{dsl} lowering in the compiler.}\label{fig:dialect}
\end{figure}

% Goals and non-goals.
As this dialect replaces the \gls{ast}, it does not perform aggressive canonicalization and instead attempts to preserve the input program as closely as possible, except for type declarations.
CFDlang's implicit scalar type greatly simplifies its type system and unclutters this particular \gls{mlir} representation.
The set of operations is also kept extremely simple, mapping language elements $1:1$ onto operations in the CFDlang dialect, with elements such as \texttt{cfdlang.eval} taking the place of identifier expressions and the like.
Transformations and optimizations are left to the middle end.

\subsubsection{The Middle-end}

% Motivation.
The program description obtained in \autoref{fig:dialect:translation} is tied to the CFDlang \gls{dsl}, which is not desirable for implementing reusable optimization pipelines.
We address this issue by introducing another level of abstraction based on the concept of tensors as immutable values, implemented by the \texttt{teil} dialect.
The exact semantics of this cross-domain tensor abstraction are directly imported from its specification~\cite{Rink2019}.
We first lower onto this dialect as illustrated by \autoref{fig:dialect:lowering}, removing \gls{dsl}-specific elements from the \gls{ir}, and then continue to transition downwards as outlined in \autoref{fig:dialectDeps}.
The set of abstractions and transformations starting with and below the \texttt{teil} dialect make up our compiler's middle-end.

% The why of TeIL.
Unlike \gls{mlir}'s \texttt{linalg} dialect for linear algebra and tensor expressions, \texttt{teil} is more restrictive on the operations it models.
Similarly, \texttt{teil} does not allow for partially defined values or incomplete domains but also prohibits treating tensors as arrays.
\texttt{teil} is a true value-based dialect with tensors as first-class citizens without any links to array materialization.
It is therefore more related to the \texttt{vtensor} concept of torch-mlir\footnote{\url{https://github.com/llvm/torch-mlir}} and the \texttt{linalg\_ex} extensions found in IREE\footnote{\url{https://github.com/google/iree}}.
This allows us to perform deductive reasoning on complex tensor operators, assigning them to array buffers later.

\texttt{teil} is another dialect bridging the gap between other high-level tensor dialects, such as the related HLO and TOSA dialects, and lower dialects, such as \texttt{linalg} and \texttt{affine}.
While HLO\footnote{\url{https://github.com/tensorflow/mlir-hlo}}, TOSA\footnote{\url{https://developer.mlplatform.org/w/tosa/}} and TPP~\cite{Georganas2021} are closer to \glspl{isa} for tensor accelerators in \gls{ml} applications, \texttt{teil} aspires to be a cross-domain tensor expression dialect.
For example, \texttt{tosa.matmul} (\autoref{fig:gemm:tosa}) encodes the well-known \gls{gemm}, which is broken down into primitive operations in \texttt{teil} as shown in \autoref{fig:gemm:teil}.
Trying to map arbitrary \texttt{teil} programs onto \texttt{tosa} or HLO is generally not possible, however, for reasons of unsupported operators and data types.

\begin{figure}[htbp]
    \centering
    \begin{minipage}{0.38\textwidth}
    \begin{subfigure}{\textwidth}
\begin{lstlisting}[language=mlir,basicstyle=\tiny\ttfamily,frame=tb,numberstyle=\tiny,linewidth=0.95\columnwidth,xleftmargin=2em,firstline=1,lastline=1]
%C       = "tosa.matmul"(%A, %B)
\end{lstlisting}
        \vspace{-10pt}\caption{\texttt{tosa}}\label{fig:gemm:tosa}
    \end{subfigure}
    \vspace{15pt}
    
    \begin{subfigure}{\textwidth}
\begin{lstlisting}[language=mlir,basicstyle=\tiny\ttfamily,frame=tb,numberstyle=\tiny,linewidth=0.95\columnwidth,xleftmargin=2em,firstline=3,lastline=5]
%AB      = teil.prod %A %B
%AB_ikkj = teil.diag 2 3 %AB
%C       = teil.red add 2 %AB_ikkj
\end{lstlisting}
        \vspace{-10pt}\caption{\texttt{teil} equivalent of (\subref{fig:gemm:tosa}) }\label{fig:gemm:teil}
    \end{subfigure}
    \end{minipage}
    \begin{subfigure}{0.52\textwidth}
\begin{lstlisting}[language=mlir,basicstyle=\tiny\ttfamily,frame=tb,numberstyle=\tiny,linewidth=0.95\columnwidth,xleftmargin=2em]
affine.parallel (%i, %j) = (0, 0) to (16, 16) {
    %1 = affine.for %k = 0 to 16 iter_args(%2 = %0) {
        %3 = affine.load %A[%i, %k]
        %4 = affine.load %B[%k, %j]
        %5 = arith.mulf %3, %4
        %6 = arith.addf %2, %5
        affine.yield %6
    }
    affine.store %1, %C[%i, %j]
}
\end{lstlisting}
        \vspace{-10pt}\caption{Simple \texttt{affine}}\label{fig:gemm:affine}
    \end{subfigure}
   \vspace{-5pt}\caption{\gls{gemm} in \gls{mlir} (types omitted)}\label{fig:gemm}
\end{figure}
    
\subsubsection{The Back-end}
Lowering \texttt{teil} to the \texttt{affine} dialect (cf. \autoref{fig:gemm:affine}), we integrate with our polyhedral optimizations for \gls{cfd} applications described in~\cite{Friebel2021}.
Starting with \texttt{affine}, the \gls{mlir} stack also allows us to import optimizations for mainstream CPUs such as in~\cite{Rink2018}.
In the next section, we will show how this maps to the template architecture from \autoref{fig:target_system}.

% Base2.
We designed the \texttt{base2} dialect to include arbitrary-precision floating-point operations.
Both \texttt{cfdlang} and \texttt{teil} use an abstract scalar type, implemented by \texttt{base2}'s parametric types, and operations that model arbitrary-precision data.
We can then use the \texttt{ieee754} type to encode custom floating-point types that can be consumed by subsequent HLS tools, like Bambu~\cite{Ferrandi2021}.

\subsubsection{Limitations}\label{sec:compiler_limitations}

We support programs that can be directly mapped onto the primitives of TeIL~\cite{Rink2019}, which is currently the most abstract middle-end representation of our compiler. On one hand, these primitives are common in many scientific applications similar to \gls{cfd} simulations. On the other hand, the \gls{mlir} ecosystem \textquote{at large} is modular and flexible, offering additional dialects that can be potentially lowered towards \texttt{affine} and \texttt{base2}, extending our flow to more domains.

A downside to using the CFDlang \gls{dsl} for this demonstration is that it does not allow for dependencies between the implicit elements.
Solvers commonly examined on FPGA platforms include Lattice-Boltzmann methods (LBM)~\cite{Sano2007}, which need to exchange halo regions with neighboring compute units.
However, for single devices, we expect this problem to be negotiated through memory, via overlapped compute and alternating kernels, thus still utilizing most of our flow.

\subsection{Compilation Process}\label{sec:compiler}

\begin{figure}
    \centering
    \includegraphics[width=0.7\columnwidth]{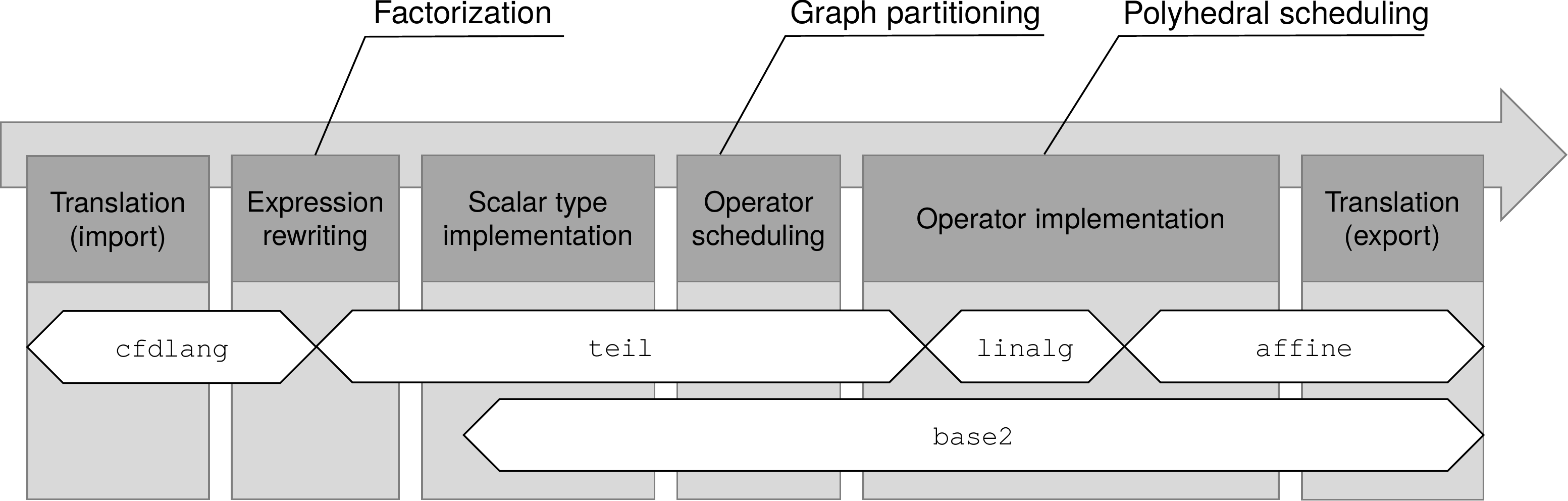}
    \caption{Steps in the CFDlang compiler, including \gls{mlir} dialects and principal transformations.}\label{fig:compilerFlow}
\end{figure}

% Introduction.
In~\cite{Friebel2021} we described analysis and transformations at the kernel level that allowed for memory allocation and operation scheduling for FPGA offloading.
These were based on polyhedral techniques, which still apply to the \texttt{affine} intermediaries that our \gls{mlir} flow produces.
For example, the liveness analysis required for Mnemosyne's sharing optimization follows the same approach described in~\cite{Friebel2021}.
To address the \gls{hbm} challenges (cf. \autoref{sec:challenges}), we introduce new transforms at higher levels, namely operator graph partitioning and pipelining.
These increase the achieved memory bandwidth  required to take advantage of the high throughput of the \gls{hbm}-based target architecture.
We elaborate on \autoref{sec:compiler_arch} by discussing the individual steps as shown in \autoref{fig:compilerFlow}.

\subsubsection{Expression Rewriting}

% Motivation.
The \texttt{teil} dialect provides a high-level view of the tensor expressions, allowing them to be rewritten effectively.
% Expression rewriting is semantically-preserving.
During the expression rewriting phase, our compiler uses strictly beneficial mathematical identities to reduce the runtime complexity of the program.
Within \texttt{teil}, which uses abstract scalars modeling $\mathbb{R}$, these are always semantically preserving.

\begin{figure}
    \centering
    \includegraphics[width=0.65\columnwidth]{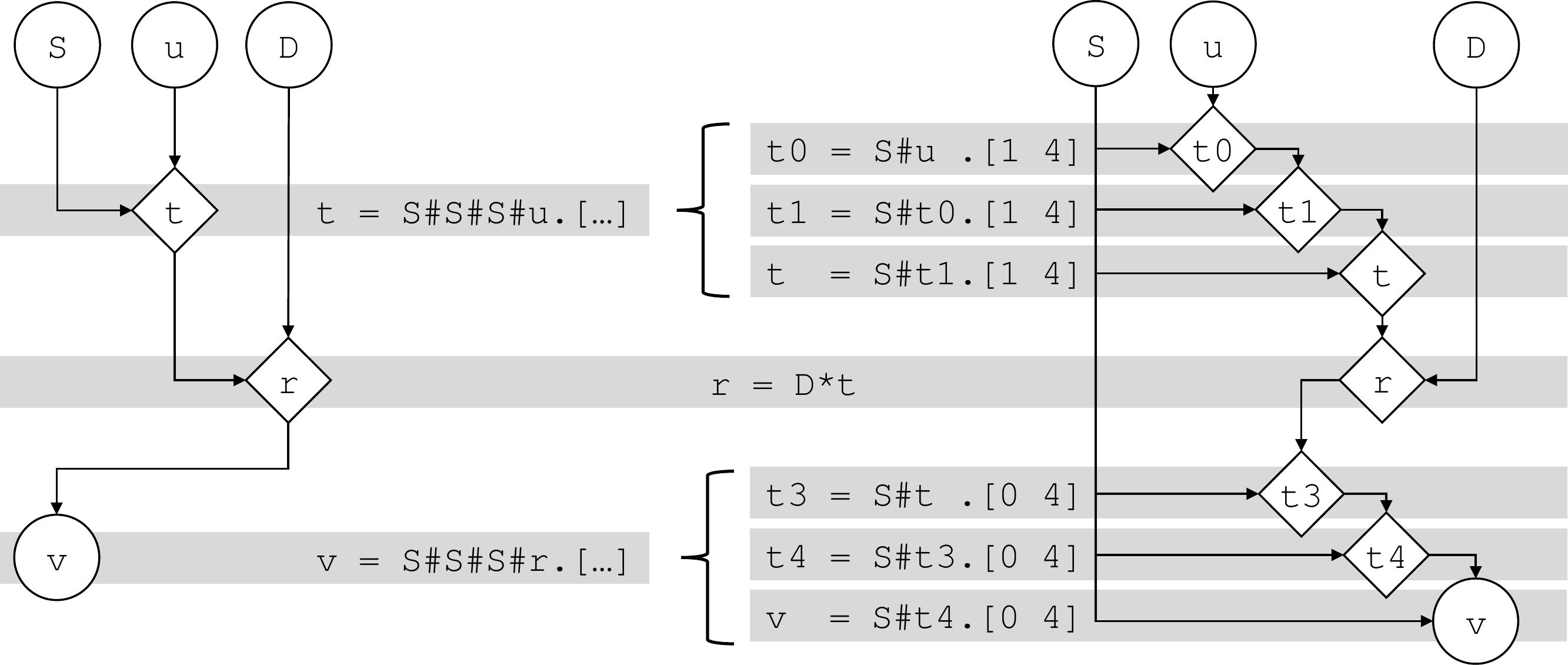}
    \vspace{-6pt}
    \caption{Expression rewriting example. Circles and diamonds indicate buffers and virtual tensors, respectively. Arrows are added to indicate dataflow dependencies, and nodes are ordered according to a naive schedule.}\label{fig:factorization}
\end{figure}

% Relation to MLIR.
One such rewrite, introduced to CFDlang in~\cite{Rink2018}, uses associativity and distributivity to factorize contractions, as shown in \autoref{fig:factorization}.
This exact rewrite can also be observed in \autoref{fig:dialect:lowering}, where the input expression is given by \autoref{fig:dialect:translation} (with a \texttt{teil} lowering beforehand).
In this transformation, the computational complexity of a contraction on an outer product is reduced by pulling it (partially) down to the factors.
The design of \texttt{teil} makes these scenarios easy to recognize and fully automates this and other kinds of rewriting, such as aggressively transforming towards \gls{gemm} patterns.

\subsubsection{Scalar Type Implementation}

% Motivation.
\Glspl{dsp}, which are used to implement IEEE-754 floating-point operations, are critical resources in FPGAs. In addition to being relatively scarce and routing intensive, when compared to \glspl{lut} and \glspl{ff}, floating-point operations require multiple clock cycles.
Such long pipelines are especially large design hazards once the design frequency drops, as the adverse effects are multiplicative.

% Proposal.
To reduce the impact of floating-point operations, our compiler infrastructure supports custom number representations.
While the existing CPU compilation flow uses \texttt{double} as the default scalar type, the FPGA flow allows arbitrary-precision types.
The exploration of this design space, however, is not automated by this work and is left up to the user.
We intend on coupling the compiler with exploration frameworks such as~\cite{Changchun2003, Chatelain2019} in the future.
Implementations use the fact that \texttt{teil} can work with multiple equivalence classes of its abstract scalar type.
This allows deferring the choice of a concrete type while still retaining the ability to reason about precision boundaries.

% Relation to Bambu.
As a starting point, we developed the \texttt{base2} dialect that provides a minimal interface to encode this information, which must be processed by a back-end consumer.
To test our
implementations on the CPU using software-emulated arbitrary-precision floating-point arithmetic, we emit calls to libsoftfloat\footnote{\url{https://github.com/ucb-bar/berkeley-softfloat-3}}.
Additionally, calls to this library can be conveniently synthesized to hardware arbitrary-precision floating-point logic by HLS tools that support this feature, like Bambu~\cite{Ferrandi2021}.

\subsubsection{Operator Scheduling}

% Motivation.
To saturate the bandwidth of the \gls{hbm} memory interface, the compiler must generate an implementation that maximizes the throughput from and into the \glspl{pc}.
This is mostly done by trading resources for frequency and cycle count using chained pipelines. 

\begin{figure}[htbp]
    \centering
    \includegraphics[width=0.65\columnwidth]{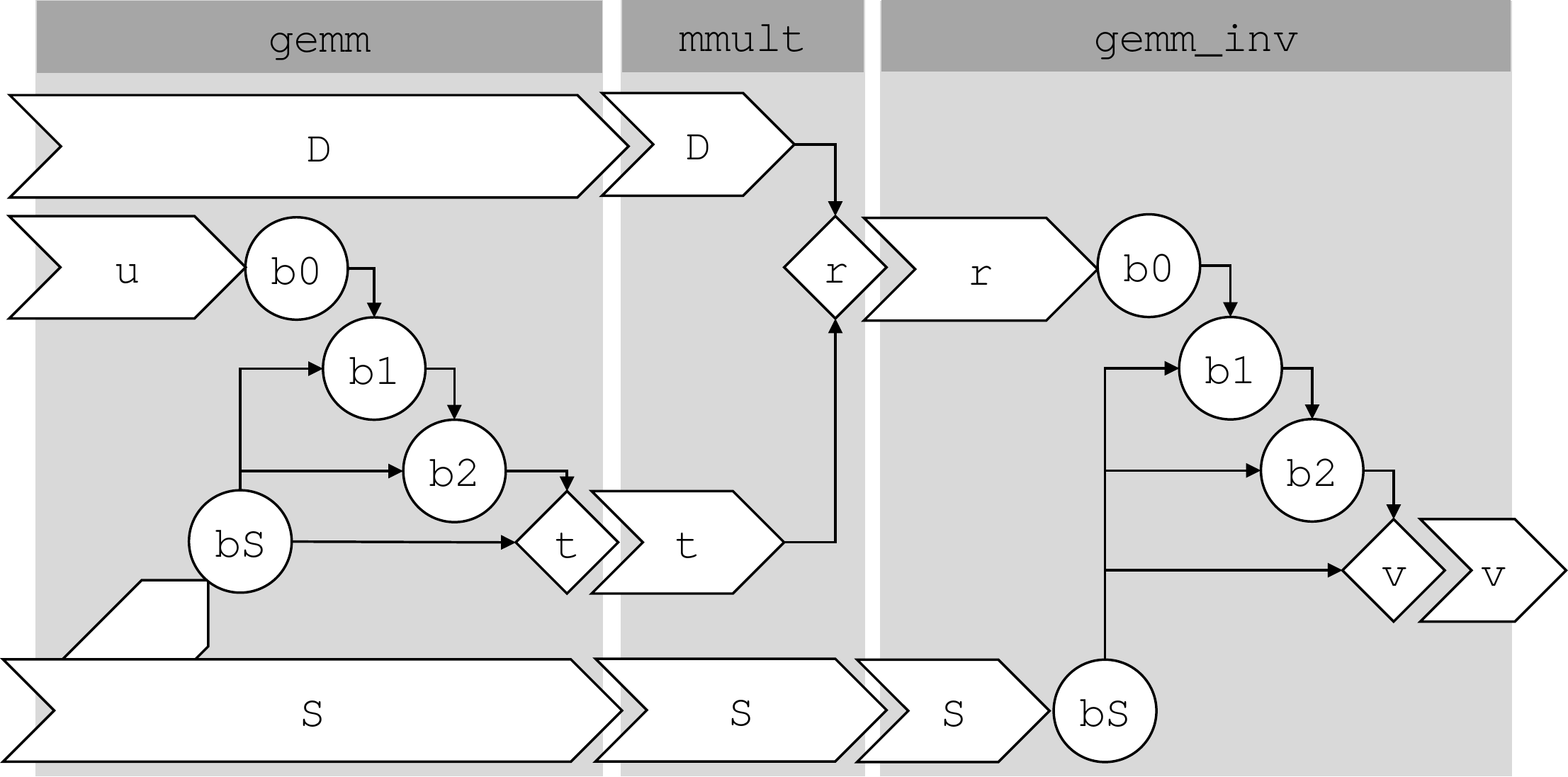}
    \caption{Operator scheduling example. The program from \autoref{fig:factorization} is partitioned into 3 groups (\texttt{gemm}, \texttt{mmult}, and \texttt{gemm\_inv}). Interfaces between stages are turned into streams, as indicated by the broad arrows.}\label{fig:operator_scheduling}
\end{figure}

% Operator graph.
Compare \autoref{fig:factorization} for the level of abstraction that this transformation is performed on, which is the tensor value graph.
Following all rewrites to the whole input program from \autoref{fig:helm_op:dsl}, this operator schedule arrives at the graph shown in \autoref{fig:operator_scheduling}.
The additions here are the operator groups, represented by grey boxes, and streams, represented by arrows.
Within each box, standard FPGA-directed pipeline scheduling will be implemented.

% Pipelining problem.
To achieve maximum throughput, a grouping must be found that allows streaming the inputs and outputs at saturating rates.
Unfortunately, although these constraints can be propagated from the \glspl{pc}, rates are not reliable measurements for this process.
The actual throughput of the hardware design depends on the achieved design frequency, which remains a variable until the end, leaving timings uncertain.
As a result, our current implementation relies either on user input or assumes that no variables outside of its control will lower the design frequency.

% Best-effort strategy.
Our current strategy first aggressively partitions the graph into the smallest possible operators (i.e.\ one per tensor value) and then collapses them.
Assuming fewer stages are better since they use fewer resources, operators can be merged automatically under a given \gls{plm} and \gls{dsp} budget.
This heuristic prefers collapsing chains, thus reducing the \gls{fifo} queues required to implement the top-level dataflow.
For our running example, this strategy yields the three top-level groups shown and named in \autoref{fig:operator_scheduling}.

    % Cycle interval budget.
Aside from the groups shown here, the template architecture fixes two groups implicitly, namely the read from and write to \gls{hbm} (cf. \autoref{fig:target_system}).
We found that our designs synthesized in a way where the group cycle intervals can be reasonably estimated by the sum of trip counts of their child loops.
When collapsing, the group with the longest interval determines the lower bound on the achievable latency, and thus our heuristic uses that interval as a budget to collapse towards.

% Pipeline scheduling.
Starting from \autoref{fig:operator_scheduling}, we can obtain an implicit top-level pipeline over the groups.
Our simple implementation selects one topological ordering, \gls{alap}~\cite{Baruch1996}, which is later used for the system generation (cf. \autoref{fig:toolflow}).
Respecting the concurrent nature of hardware implementations, future extensions may consider merging stages that can execute in parallel.

\subsubsection{Operator Implementation}

% Motivation.
To ensure the group pipeline scheduled in the previous step runs as designed, the inner nodes of each group must be implemented as a regular pipelined loop nest.

% Previous extensions.
In~\cite{Friebel2021}, we extended the CFDlang compiler with polyhedral analysis and code generation capabilities.
This framework allowed us to implement transformations to render the code amenable to further processing with \gls{hls} tools.
In particular, we incorporated changes to the memory layout and loop structure, alleviating resource burdens and improving pipelining.
Here, the polyhedral model was an especially good fit for liveness analysis, which allowed us to defer an implementation problem, the memory sharing, to the C-to-system generation part.

% Putting it together.
We lower from \texttt{teil} to \texttt{affine} and plumb directly into this previous code generator from~\cite{Friebel2021} (cf. \autoref{fig:cont_1}).
This also allows us to reuse code generation via the \gls{isl}~\cite{Verdoolaege2010}, circumventing the need for a new \gls{hls} back-end dialect.
% Concrete implementation problems.
The streaming property of tensors between groups can be trivially upheld using polyhedral scheduling by constraining the order of the writes.
More than one output stream per operator requires an additional effort that was not needed for this work but has been examined previously~\cite{Verdoolaege2017}.
A stopgap solution lies in buffering the reads in the groups (as shown in \autoref{fig:operator_scheduling}), which may sometimes even be a necessary resource trade-off if consecutivity cannot be established.
In~\cite{Friebel2021} we described memory sharing optimizations that reduce this pressure on the FPGA's BRAM again.
For the kernel evaluated in this paper, memory sharing was not as relevant, especially when increasing the number of groups. In fact, a downside of stage level sharing is that it may limit the ability to pipeline the internals of the stage, ultimately sacrificing throughput and increasing the cycle interval.

\begin{figure}[htbp]
    \centering
    \begin{subfigure}{0.9\textwidth}
\begin{lstlisting}[language=mlir,basicstyle=\tiny\ttfamily,frame=tb,numberstyle=\tiny,linewidth=0.95\columnwidth,xleftmargin=2em,breaklines=true,postbreak=\mbox{\textcolor{red}{$\hookrightarrow$}\space}]
affine.for %arg5 = 0 to 11 {
    affine.for %arg6 = 0 to 11 {
        affine.for %arg7 = 0 to 11 {
            affine.for %arg8 = 0 to 11 {
                %6 = affine.load %bS[%arg5, %arg8] : memref<11x11xf64>
                %7 = affine.load %b0[%arg6, %arg7, %arg8] : memref<11x11x11xf64>
                %8 = arith.mulf %6, %7 : f64
                %9 = affine.load %b1[%arg5, %arg6, %arg7] : memref<11x11x11xf64>
                %10 = arith.addf %8, %9 : f64
                affine.store %10, %b1[%arg5, %arg6, %arg7] : memref<11x11x11xf64>
            }
        }
    }
}
\end{lstlisting}
        \vspace{-10pt}\caption{in \texttt{affine}\vspace{6pt}}
    \end{subfigure}
    \begin{subfigure}{0.9\textwidth}
\begin{lstlisting}[language=c,basicstyle=\tiny\ttfamily,frame=tb,numberstyle=\tiny,linewidth=0.95\columnwidth,xleftmargin=2em,breaklines=true,postbreak=\mbox{\textcolor{red}{$\hookrightarrow$}\space}]
for (int c1 = 0; c1 <= 10; c1 += 1)
    for (int c2 = 0; c2 <= 10; c2 += 1)
      for (int c3 = 0; c3 <= 10; c3 += 1) {
        // stmt0
        b1[121 * c1 + 11 * c2 + c3] = 0;
        for (int c4 = 0; c4 <= 10; c4 += 1) {
            #pragma HLS pipeline
            // stmt0
            b1[121 * c1 + 11 * c2 + c3] = b1[121 * c1 + 11 * c2 + c3] + bS[11 * c1 + c4] * b0[121 * c2 + 11 * c3 + c4];
        }
      }
\end{lstlisting}
        \vspace{-10pt}\caption{in C99}
    \end{subfigure}
   \vspace{-5pt}\caption{\autoref{fig:dialect:lowering} lines 1-3 during operator implementation}\label{fig:cont_1}
\end{figure}

\subsubsection{Automation}\label{sec:automation}

There are two sides to automation of these transforms in the \gls{mlir} flow because of the separation of policy and mechanism.
The \texttt{teil} dialect allows us to fully automate tensor expression rewriting, within the confines of mathematical reasoning, including decision making.
The \texttt{base2} dialect in conjunction with \texttt{teil} provides a mechanism for arbitrary-precision scalars but offers no support for exploring that design space.
In \texttt{teil}, we can encode partitions of operator pipelines with ease, but as we also see in vendor toolchains, simple heuristics can not replace proper design space exploration based on performance estimation in the general case.
The \gls{isl} is a powerful tool that can make and enforce decisions (e.g.\ schedule loops based on streaming constraints), but has an inherently biased and non-deterministic scheduler.

For custom data types, we currently rely on command-line parameters to guide the grouping in the presence of design frequency hazards and apply array layout and partitioning maps~\cite{Friebel2021}. 
While we aim at keeping these hints out of the \gls{dsl}, they can be represented using meta-operations in the \gls{ir}. 
An example of this is HeteroCL~\cite{Lai2019}, and this has already been proposed for TeIL~\cite{Susungi2018}.

\subsection{Hardware Generation Architecture}\label{sec:mem}

Our hardware generation flow aims at optimizing the data transfers around the kernel implementation produced by CFDlang.
It receives as input the description of the kernel and the compatibility graph of the internal buffers (from CFDlang), along with information about board resources (from the user). It produces an optimized \gls{cu} description (in C++) and the platform configuration file based on the number of \glspl{cu} that can be instantiated. The configuration file also specifies the proper connections to the \gls{hbm} channels.
To implement our hardware generation flow, we developed Olympus, which creates an optimized system-level architecture (in synthesizable C++) for our accelerators. During its execution, it interfaces with Mnemosyne~\cite{Pilato2017} to optimize the on-chip memory associated with each kernel and limit the number of local memory resources. 
Mnemosyne uses the buffer compatibility graph generated by the CFDlang compiler to determine when the physical on-chip memory banks can be reused without performance overhead~\cite{Pilato2017}. After this, Olympus reads the kernel interface and determines how to connect the input/output ports to the rest of the system to efficiently exchange data with the \gls{hbm} channels. Data ports are connected to \gls{hbm} channels via AXI Master interfaces, while configuration ports are connected to the host via AXI-Lite, memory-mapped interfaces. Data exchanged with \gls{hbm} channels (i.e., input and output matrices) are buffered on-chip to allow fast, fixed-latency access during kernel execution.

\begin{figure}[tbp]
\centering
    \begin{subfigure}{0.95\columnwidth}
        \centering
\begin{lstlisting}[language=C,basicstyle=\tiny\ttfamily,frame=tb,linewidth=\columnwidth,xleftmargin=2em,framexleftmargin=1.5em]
void kernel(double S[121], double D[1331], double u[1331], double v[1331], double t0[1331], 
            double t1[1331], double t[1331], double r[1331], double v0[1331], double v1[1331]) {
    //...
}
\end{lstlisting}
        \vspace{-10pt}\caption{After CFDlang\vspace{6pt}}
    \end{subfigure}
    \begin{subfigure}{0.95\columnwidth}
        \centering
\begin{lstlisting}[language=C,basicstyle=\tiny\ttfamily,frame=tb,linewidth=\columnwidth,xleftmargin=2em,framexleftmargin=1.5em]
void top_kernel(double S[121], double D[1331], double u[1331], double v[1331]) {
    double t0[1331], t1[1331], t[1331], r[1331], v0[1331], v1[1331];
    kernel(S, D, u, v, t0, t1, t, r, v0, v1);
}
\end{lstlisting}
        \vspace{-10pt}\caption{After Mnemosyne\vspace{6pt}}
    \end{subfigure}
    \begin{subfigure}{0.95\columnwidth}
        \centering
\begin{lstlisting}[language=C,basicstyle=\tiny\ttfamily,frame=tb,linewidth=\columnwidth,xleftmargin=2em,framexleftmargin=1.5em]
void top_kernel(double S[121], double D[1331], double u[1331], double v[1331]) {
    double t0[1331], t1[1331];
    kernel(S, D, u, v, t0, t1, t0, t1, t0, t1);
}
\end{lstlisting}
        \vspace{-10pt}\caption{After Mnemosyne with array sharing}
    \end{subfigure}
\caption{Kernel interface before and after Mnemosyne. Internal buffers are exposed by CFDlang. Mnemosyne reproduces a kernel description with only \textquote{real} input and output buffers for interfacing with the \gls{hbm} channels. Other temporary on-chip buffers are internally optimized and possibly shared.}
\label{fig:mnemosyne}
\end{figure}

The generation of the system architecture is guided by the designer, as shown in \autoref{fig:toolflow}. Starting from the C kernel produced by the CFDlang compiler, Olympus generates a minimal wrapper to run HLS and obtain an estimate of the resources needed for the target FPGA device.
Then, we apply optimizations to both the on-chip local memory of each kernel (with Mnemosyne) and the system-level memory architecture to exchange data with the \gls{hbm} (with Olympus).

At the kernel level, we use Mnemosyne to generate the RTL of the on-chip memory architecture associated with each kernel. Mnemosyne is a tool that exploits sharing compatibilities, i.e., when distinct internal buffers have no overlapping lifetime, to assign them to the same physical banks based on a given cost metric. In these cases, the tool generates custom logic to manage the accesses to the same memory banks from different kernel interfaces~\cite{Pilato2017,Friebel2021}. \autoref{fig:mnemosyne} shows the conceptual interface, described in C, of a kernel before and after the execution of Mnemosyne. Mnemosyne requires the specification of the compatibility graph of the local arrays and, in our flow, these metadata are computed and produced by CFDlang during liveness analysis~\cite{Friebel2021}. Mnemosyne wraps the RTL kernel description (produced by HLS) with the resulting RTL description of the kernel memory architecture to expose only input and output ports to the \gls{cu}. This conceptual interface is then used to integrate the kernel into the \gls{cu}.

At the system level, Olympus generates the C++ description of the memory architecture around the kernel description, which is integrated as custom RTL to use the results produced by Mnemosyne.
Since the hardware cost of the kernel may limit the number of parallel \glspl{cu}, the designer can use Olympus to understand which optimizations can be applied given the available FPGA resources. Indeed, we characterize each optimization with an estimation of the extra resources. With this information, the designer can select the most suitable optimizations, and Olympus generates the corresponding \gls{cu} description around the CFDlang-generated code of the kernel and the system configuration file. In the future, this process can be further automated by combining it with state-of-the-art design space exploration frameworks.
Each \gls{cu} can feature multiple kernels, each of them connected to a lane to fully utilize the AXI bandwidth (cf. \autoref{sec:system}). The \gls{cu} wrapper implements data-movement optimizations and is designed accordingly with changes to the host application and the configuration file.
For example, the kernel may benefit from a change in the way data is written to and read from global memory and therefore the host application must be updated accordingly.
The configuration file, instead, defines how each \gls{cu} interfaces with the \gls{hbm}. By modifying this file, Olympus defines how each \gls{cu} is connected to the individual channels.

The resulting components are then passed to Vitis, i.e., the \gls{hls} platform that we use, to automatically generate the bitstream required for board configuration. When timing is not met, Vitis automatically downscales the execution frequency. While this is useful to enable proper functionality, the designer has little control over this process. Conversely, Olympus introduces some optimizations for the synthesis process (cf. \autoref{sec:res_opt}). 

\subsubsection{Limitations}\label{sec:mem_limitations}
Our hardware generation flow does not currently include a step for including platform-specific optimizations, like the mapping of array streams to specific memory resources. This step also needs customization when bringing the flow to a new \gls{hls} tool or platform as it requires the insertion of the proper directives.

Also, our flow must emit C99 source code to interface with the given \gls{hls} tool. A direct interface would be beneficial to avoid losing semantic details that are contained in the MLIR dialects.

\subsection{Hardware Generation Process}\label{sec:mem_opt}

Let $p$ be the polynomial degree for the simulation, each Inverse Helmholtz kernel needs $p\times p$ values for matrix \texttt{S} and $p\times p \times p$ values for matrices \texttt{D} and \texttt{u}. The kernel then produces  $p\times p \times p$ values for the output matrix \texttt{v}. These values are useful to estimate the number of resources for input, temporary, and output buffers. 
The size of the total amount of data used to compute one element can be used to determine how many elements worth of data can fit into an \gls{hbm} channel (max size is 256MB). The number of elements is the \textbf{batch size}, i.e., the number of executions the \gls{cu} can perform without interruption before needing the host to send more data.
Data transfers are required between consecutive batches.

In the following, we describe the optimizations that we apply to our \gls{cfd} application by means of the Olympus options (step \textit{Optimize} in \autoref{fig:toolflow}). Such optimizations (and the Olympus hardware generation process) can be easily adapted to similar tensor-based applications.

\begin{figure}
     \begin{subfigure}[b]{0.48\columnwidth}
         \centering
         \includegraphics[width=\textwidth]{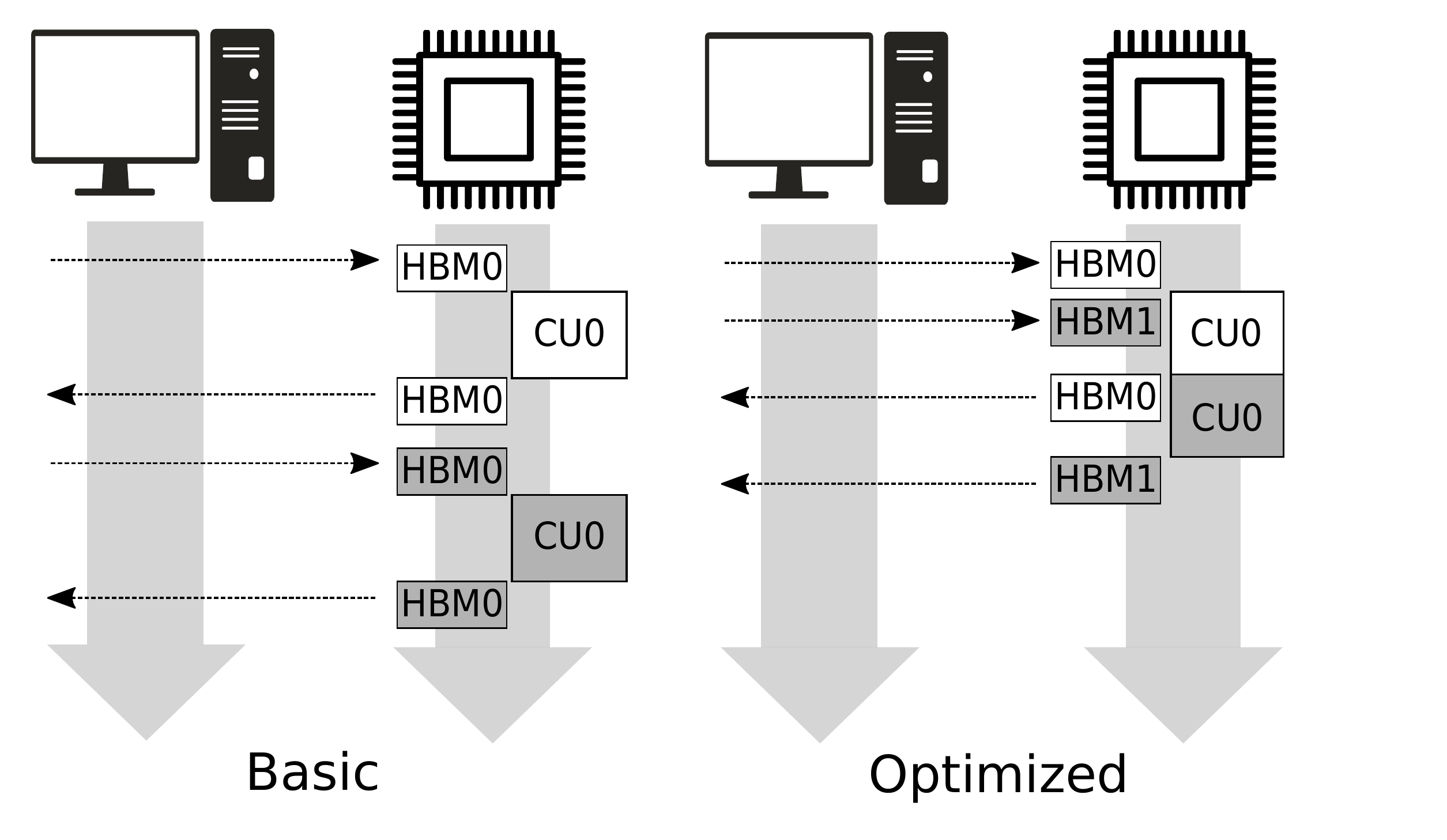}
         \caption{Host-HBM Double Buffering}
         \label{fig:hbmpp}
     \end{subfigure}
     \hfill
     \begin{subfigure}[b]{0.48\columnwidth}
         \centering
          \includegraphics[width=\textwidth]{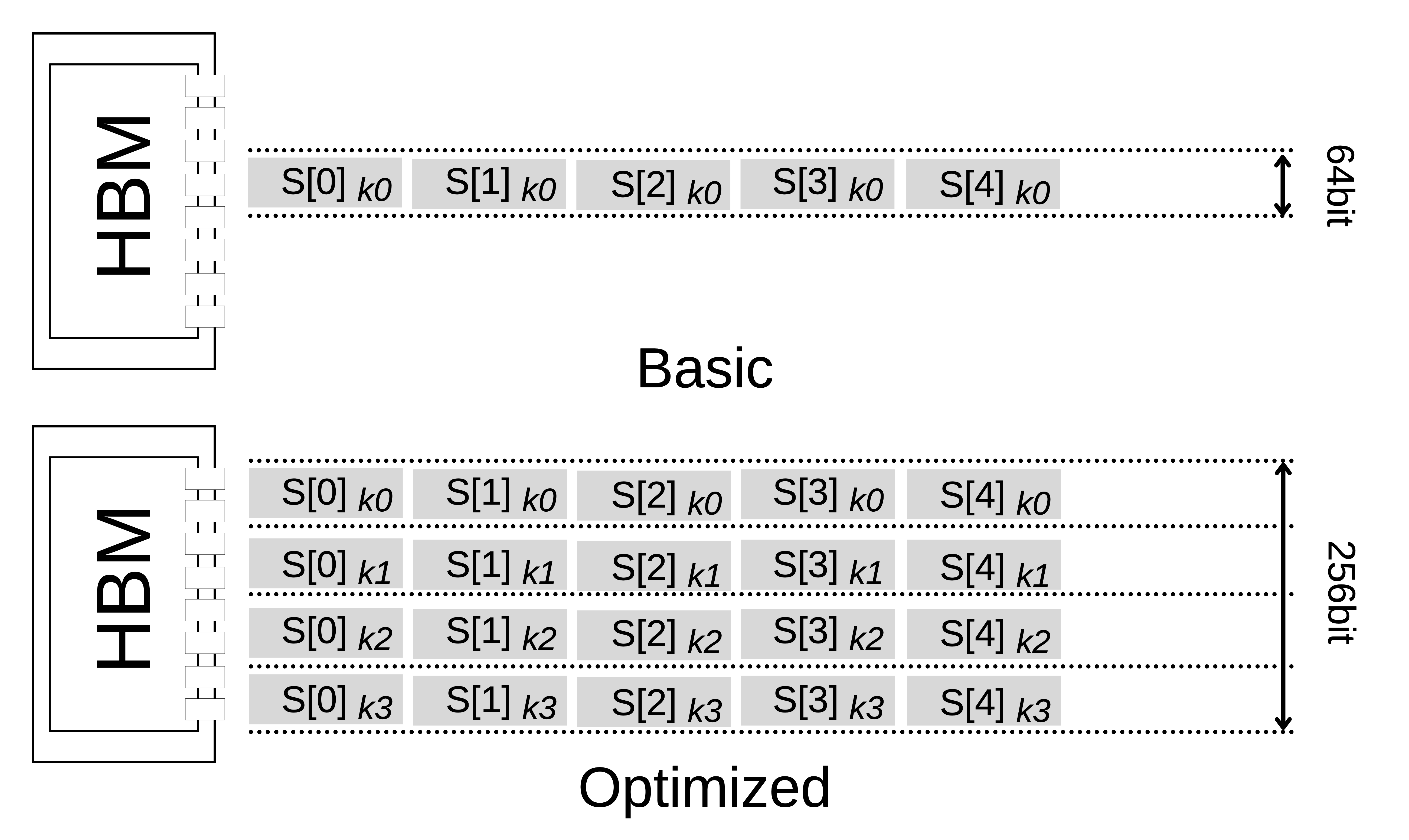}
         \caption{Bus Optimizations}\vspace{6pt}
         \label{fig:bus}
     \end{subfigure}
     \begin{subfigure}[b]{0.48\columnwidth}
         \centering
          \includegraphics[width=\textwidth]{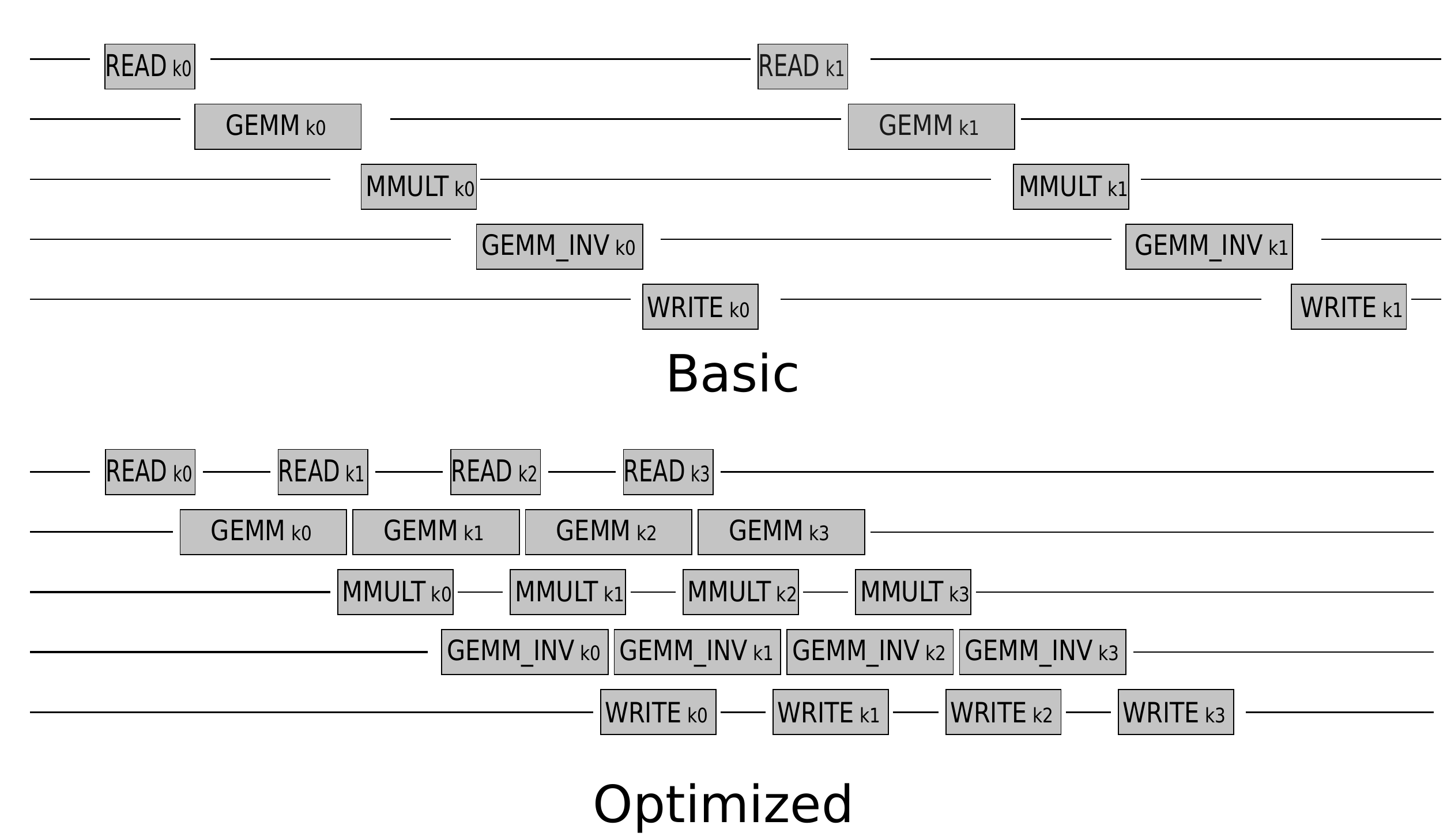}
         \caption{Dataflow Optimization}
         \label{fig:interleaving}
     \end{subfigure}
     \hfill
     \begin{subfigure}[b]{0.48\columnwidth}
         \centering
         \includegraphics[width=0.9\textwidth]{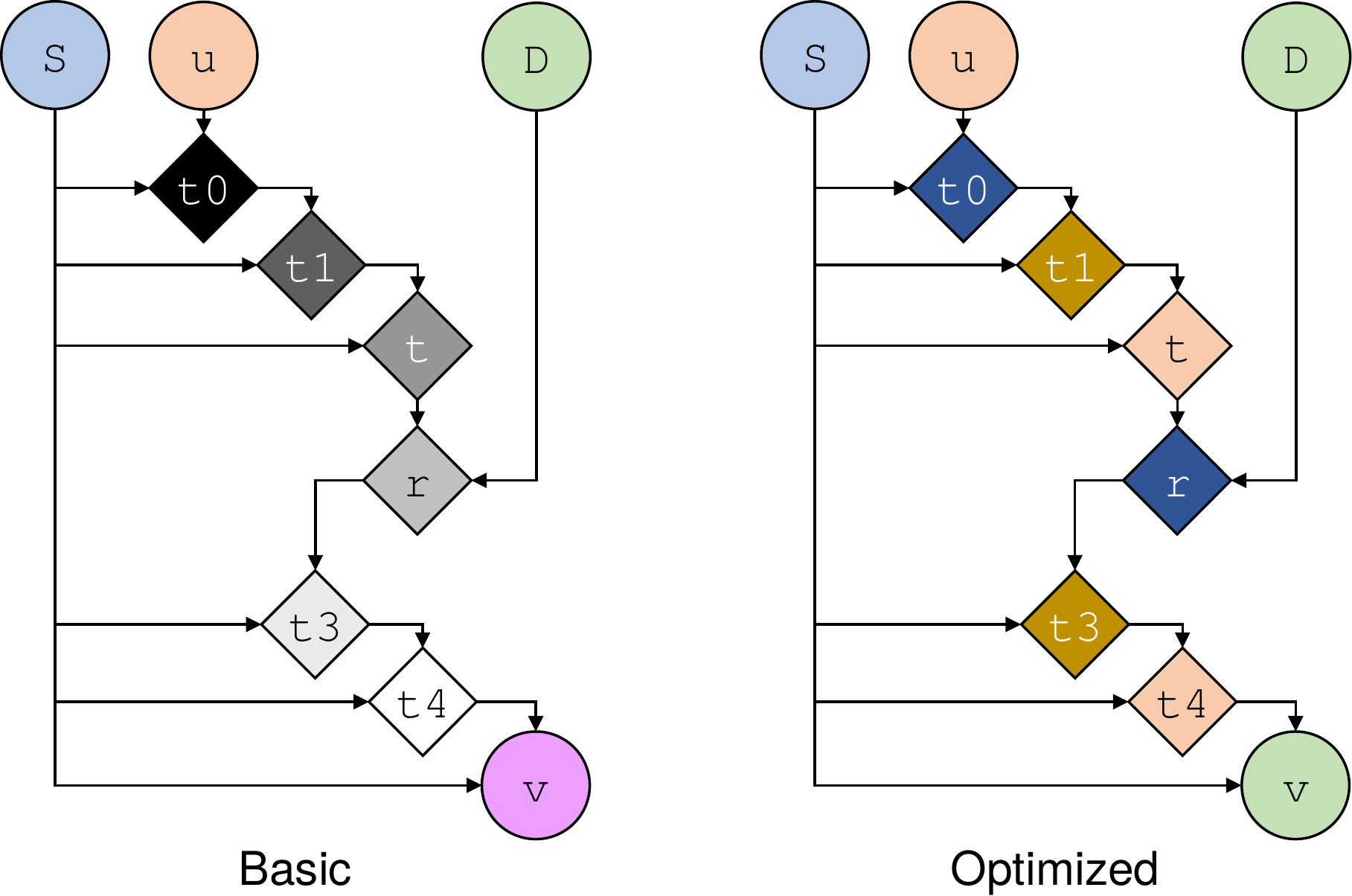}
         \caption{Resource Optimization}
         \label{fig:sharing}
     \end{subfigure}
\caption{\vspace{-9pt}Optimizations targeting the memory and communication challenges of the \gls{hbm}-based systems.}\label{fig:opts}
\end{figure}

\subsubsection{Host-\glsentryshort{hbm} Double Buffering}
In a naive implementation, the host code transfers the input data required to execute one batch of kernel elements into \gls{hbm}. The host then invokes the \glspl{cu} to execute on each of these elements and generate the corresponding output results. The host transfers these outputs back from \gls{hbm} to its main memory. Each \gls{cu} interfaces with one \gls{pc} and we can instantiate up to 32 \glspl{cu} (each with one kernel) to operate in parallel. However, all communication and execution for a single \glspl{cu} is serialized. Since the host-\gls{hbm} communication is as expensive as the computational part, this significantly affects the overall performance.

To overlap the host-\gls{hbm} data transfers with the \gls{cu} execution, we use double buffering as shown in \autoref{fig:hbmpp}. Each \gls{cu} interfaces with two \glspl{pc}, namely \textquote{even} and \textquote{odd}. The host reads the output from the last iteration and writes new input into the \textquote{even} channels while the \glspl{pc} operate on the data in the \textquote{odd} channels, and vice versa.
When the total host transfer time for input and output of one batch is less than the total \gls{cu} execution time for the same batch, the host transfer time is entirely hidden, and the \glspl{cu} are actively executing at all times. 

\textsc{Olympus Implementation:} Double buffering 
requires changes in the wrapper to determine which \gls{pc} the \gls{cu} should operate on for the current batch, in the configuration file to attach more channels to the same \gls{cu}, and in the host code to manage the data transfers to separate \gls{hbm} channels. Additionally, since we use two \glspl{pc} to implement double buffering, we limit the maximum number of parallel \glspl{cu} to 16. However, when the total number of \glspl{cu} that can be instantiated is less than 8, we also separate input and output channels to simplify the control logic and improve logic connectivity of the FPGA resources.

\subsubsection{\gls{hbm}-FPGA Bandwidth Optimization}\label{sec:bus_opt} 

The data elements of the Inverse Helmholtz operator are 2- or 3-D tensors ($p\times p$ or $p\times p\times p$, respectively), where each element is a 64-bit floating-point number (\textit{double}).
The \gls{hbm} interfaces have 256 bits, so transferring one double at a time uses only 25\% of the bandwidth. Packing four doubles (or more depending on the custom data type) onto the bus allows us to reduce the number of clock cycles for data transfers significantly. However, we need to efficiently manage the multiple parallel data to avoid serialization when writing them into the buffers.
To fully exploit the parallelism, we conceptually divide the 256-bit bus into four 64-bit ``lanes'' and replicate the kernel four times within a \gls{cu}. Each of these kernels is directly connected to one of the ``lanes'', i.e., to 64 bits of the bus, so that all kernels can operate in parallel. In this way, \textit{Read}/\textit{Write} modules still require the same amount of cycles (cf. \autoref{fig:bus}) but we can now start the computation of four elements in parallel.

\textsc{Olympus Implementation:} To obtain the layout shown in \autoref{fig:bus}, Olympus modifies the host code to interleave the input for the multiple elements before sending it to \gls{hbm} and de-interleave the output after receiving the results. The optimization only needs information on the bus bitwidth (e.g., 256 bits) and the data type bitwidth (i.e., 32 or 64 bits). Both parameters are available from the user-supplied board specification and the compiler-supplied array information, respectively. From this, Olympus generates the \gls{cu} \textit{Read} and \textit{Write} functions to split and aggregate the data into the appropriate number of lanes. The overall \gls{cu} structure is then created by composing the \textit{Read}/\textit{Write} functions with multiple instances of the kernels.
Similarly, the data reorganization portion of the host code can be generated with the same information by specializing the allocation functions of the host application.

\subsubsection{Dataflow Optimization}\label{sec:opt_dataflow}

Every single execution of the \gls{cu} in the batch must read data from the \gls{hbm}, execute the operator on them, and write back the result into the \gls{hbm}.
The Inverse Helmholtz operator is implemented in DSL as three operations: a tensor contraction, a Hadamard product, and another tensor contraction. We can further decompose the single operations into elementary blocks, as shown in \autoref{fig:operator_scheduling}.
The fundamental blocks can be implemented as subfunctions in the kernel that communicate via AXI Stream in a dataflow model. These hardware modules will thus execute in a pipelined manner, significantly improving the throughput. The number of elementary blocks in each subfunction is a tradeoff between latency and resource requirements. Indeed, having more blocks in the same subfunction increase resource sharing opportunities but also increases the latency of the pipeline stages, reducing the throughput. This optimization improves the performance but also increases the resource usage, potentially limiting the number of \glspl{cu} that can be instantiated.

\textsc{Olympus Implementation:} The dataflow optimization is enabled by the compiler generating a kernel composed of subfunctions using streams instead of one flat kernel function. The exact scheduling of the stages may not be as straightforward as the three conceptual operators, as the compiler has the freedom to optimize the grouping for the best performance.
Olympus then creates data streams among the subkernels for data communication. To stream data between the subkernels, data must be buffered when the subkernel does not operate on it in the same order that it is streamed or when the same values are reused multiple times inside the same subfunction.
In most cases, this means that data streamed in gets stored in an internal buffer, then the data can be operated on using random access, and as each result is computed, it is streamed out. 
Data structures that are reused across multiple blocks (like matrix \texttt{S}) must be streamed through these blocks and buffered inside them to keep a consistent structure and avoid multiple hardware modules accessing the same data concurrently.
This optimization does not require any changes in the host code and the size of the streams can be configured by the designer when selecting the optimization. When no information is provided, the tool assumes to use the full size of the array as the size of the corresponding stream.

\subsubsection{Resource Optimizations and Multiple Compute Units}\label{sec:res_opt}
The Inverse Helmholtz operator is composed of seven loops that are executed in sequence. Each loop produces a matrix. Intermediate matrices are used by the next loops. Each of these matrices requires on-chip resources (generally BRAMs) to store the values. The number of available BRAMs can limit the number of FPGA kernels. However, once the matrix is not used anymore, the corresponding BRAM resources can be used by the same kernel to store new data.

Using the liveness information generated by the CFDlang compiler, we can reduce the number of on-chip resources required by each kernel. Arrays with disjoint lifetimes can use the same physical memory buffer as shown in~\autoref{fig:sharing}. Reducing the kernel's BRAM requirements can increase the total number of kernels that we can instantiate.
However, sharing opportunities can operate only inside each subkernel. So, the effects of this optimization may be limited. It is worth noting that we currently optimize the use of memory resources only by exploiting sharing opportunities while platform-specific optimizations (like the implementation of arrays with specific memory resources) can be integrated as an additional step.

Also, given the physical nature of the input data, we can adapt the measurement scale so that the values are always in a range between -1 and 1. This observation allows us to change how we interpret the data, passing from a floating-point representation to a fixed-point one. In particular, we use a 64-bit representation where 24 bits are assigned to the integer part (including the sign) and 40 to the decimal part. Specifically, we used the \texttt{ap\_fixed} library to specify these formats so that they can be automatically synthesized by Vitis HLS. So, these optimizations are compatible with any HLS tool (like Bambu~\cite{Ferrandi2021}) that can synthesize these formats.
This step brings with it considerable advantages. Fixed-point operations require simpler hardware than floating-point operations. Tensor operators make heavy use of multipliers. So, fixed-point operations allow designers to obtain faster hardware that uses fewer resources and consumes less energy.

In \gls{cfd}, another method to decrease resource utilization is to reduce the degree $p$. This will be at some cost to the convergence of the overall simulation, but this choice is up to the designer who provides the required value of $p$ for the simulation in the input DSL. Once this value is decided, our flow will be able to automatically instantiate more parallel compute units due to the smaller data size and reduced number of the operations/loop iterations.

\textsc{Olympus Implementation}: For buffer sharing, as discussed above, we use Mnemosyne to automatically generate optimized \gls{plm} units that can share physical banks in a way that is completely transparent to the kernel execution~\cite{Pilato2017,Friebel2021}. Olympus only combines the resulting kernel description with the rest of the system in a transparent way.
Data representation is left as a design choice for the application developer, as the tolerable error depends heavily on the application. Using this choice as input, the data type can be automatically changed in the implementation to be able to observe the effects on area, power, and performance. Fixed-point implementations only require a redefinition of the data types before \gls{hls} using the given arbitrary-precision libraries. We decided to implement the conversion from/to double in the host code to save hardware resources. However, this requires an adaptation of the data allocation functions, which receive the input values in double but need to write fixed-point values in the FPGA buffers, and the functions to retrieve the results that must implement the opposite conversion.
Finally, the polynomial degree $p$ is an intrinsic parameter of the input DSL code (cf. \autoref{fig:helm_op:dsl}). We can re-run the complete flow to generate an accelerator with a different polynomial degree to enable the proper compiler and hardware generation optimizations.

\section{Evaluation}\label{sec:results}
    
In this section, we use our DSL-to-bitstream flow to evaluate and compare several implementations of the \gls{cfd} application. To do so, we present our experimental setup (\autoref{sec:setup}), we analyze the impact of each optimization for the Inverse Helmholtz kernel (\autoref{sec:tests}), and we compare our final results with software implementations (\autoref{sec:sw_impl}).

\subsection{Experimental Setup}\label{sec:setup}

To evaluate our DSL-to-bitstream flow, we extended the flow presented in~\cite{Friebel2021} as described in \autoref{fig:toolflow}: CFDlang is implemented on top of the MLIR infrastructure,
Mnemosyne~\cite{Pilato2017} is an open-source tool\footnote{\url{https://github.com/chrpilat/mnemosyne}}, and Olympus is a new in-house prototype. Olympus is built in Python on top of the Pyverilog library~\cite{Pyverilog2015} for hardware generation (i.e., the generation of the kernel wrappers around Mnemosyne artifacts) and the Pycparser library\footnote{\url{https://github.com/eliben/pycparser}} for code generation.
With our novel flow, we targeted a Xilinx Alveo U280 card on an \amd~\cite{amdepyc} server running Centos~7. We used Xilinx Vitis 2021.1~\cite{vitis} for synthesis and bitstream creation.
Unless otherwise specified, we target a synthesis frequency of 450~MHz for both the platform and the \gls{cu} description. For each implementation, we evaluate performance and energy efficiency by using the GFLOPS and GFLOPS/W metrics, respectively. The GFLOPS metric is obtained by dividing the total number of floating-point operations executed by the application by the total execution time, while the GFLOPS/W metric is obtained by dividing the GFLOPS metric by the average power consumption of the system. To get accurate information about FPGA power consumption, we profile the power consumption during the system execution with Xilinx XRT, and we use the average value.

\subsection{Impact of Hardware Optimizations}\label{sec:tests}

\pgfplotsset{select coords between index/.style 2 args={
    x filter/.code={
        \ifnum\coordindex<#1\def\pgfmathresult{}\fi
        \ifnum\coordindex>#2\def\pgfmathresult{}\fi
    }
}}
\pgfplotstableset{%
    every head row/.style={before row=\toprule,after row=\midrule},
    every last row/.style ={after row=\bottomrule},
    every even row/.style={before row={\rowcolor[gray]{0.9}}},
    %every column/.style={precision=3},
}
\pgfplotstableread[col sep=comma,header=has colnames]{images/TRETS_data_fixed_again.csv}\pelevencuonetable
\pgfplotstableread[col sep=comma,header=has colnames]{images/addl_kernels.csv}\addlkernelstable

To evaluate the cumulative benefits introduced by each hardware optimization, we progressively apply them to the \textit{Inverse Helmholtz} operator used as a case study and so extensively discussed in this paper.
We evaluated the implementations for two polynomial degrees: $p=7$ and $p=11$. In particular, given the polynomial degree $p$, we assume that each contraction is composed of three loop nests that execute two floating-point operations (one addition and one multiplication) for $p\times p \times p~ \times~p$ times each. Similarly, the Hadamard product requires $p\times p \times p$ multiplications. So, the entire Inverse Helmholtz operator the following number of floating-point operations:
\begin{equation}\label{eq:numpoints}
N^{el}_{op} = 2\cdot[2\cdot(p\cdot p\cdot p~\cdot~p) + 2\cdot(p\cdot p\cdot p~\cdot~p) + 2\cdot(p\cdot p\cdot p~\cdot~p)]+(p\cdot p\cdot p)=(12\cdot p+1)\cdot(p\cdot p\cdot p) \end{equation} 
So, a single element is required to execute $N^{el}_{op}$=177,023 floating-point operations when $p=11$ and $N^{el}_{op}$=29,155 floating-point operations when $p=7$. The total number of floating-point operations for a \gls{cfd} simulation can be obtained as:
\begin{equation}
N_{op} = N_{eq} \times N^{el}_{op}
\end{equation}
We executed all experiments with $N_{eq}=2,000,000$, i.e., we simulated 2,000,000 elements.

\begin{figure}[t]
    \centering
\begin{tikzpicture}
    \begin{axis}[
    symbolic x coords=
        {
            Baseline,
            Double Buffering,
            Bus Opt (Serial),
            Bus Opt (Parallel), 
            Dataflow (1 compute),
            Mem Sharing (1 compute),
            Dataflow (2 compute),
            Dataflow (3 compute),
            Dataflow (7 compute),
            Fixed Point 64,
            Fixed Point 32,
            Dataflow (7 compute) - Far,
            Double p11,
            Double p7,
            Fixed Point 64 p11,
            Fixed Point 64 p7,
            Fixed Point 32 p11,
            Fixed Point 32 p7,
            Double p11cu2,
            Fixed Point 64 p11cu2,
            Fixed Point 32 p11cu3,
            Double p7cu3,
            Fixed Point 64 p7cu2,
            Fixed Point 32 p7cu4,
        },
            ylabel={GFLOPS},
            ymin=0,
            ymax=150,
            x tick label style={rotate=30,anchor=east,font=\footnotesize},
            every tick label/.append style={font=\footnotesize},
            xtick=data,
            legend style={ font=\footnotesize,at={(0.02,0.98)},anchor=north west},
            width=1\columnwidth, height=0.4\textwidth,
            ybar,
            bar width=8pt,
            nodes near coords,
            nodes near coords style={rotate=90,anchor=west,font=\footnotesize,/pgf/number format/fixed, /pgf/number format/precision=3, /pgf/number format/zerofill},
            bar shift auto=3pt,
        ]
        \addplot[black,
                x filter/.code={
        \ifnum\coordindex=9\def\pgfmathresult{}\fi
        \ifnum\coordindex=11\def\pgfmathresult{}\fi
        \ifnum\coordindex>12\def\pgfmathresult{}\fi
        },
       postaction={pattern=north east lines,pattern color=black!80}] table[x=test, y={k_gflops}] {\pelevencuonetable};
        \addlegendentry{CU}
        \addplot[black!25!teal,
                x filter/.code={
        \ifnum\coordindex=9\def\pgfmathresult{}\fi
        \ifnum\coordindex=11\def\pgfmathresult{}\fi
        \ifnum\coordindex>12\def\pgfmathresult{}\fi
        },
        ,fill=teal!20] table[x=test, y={sys_gflops}] {\pelevencuonetable};
        \coordinate (A) at (axis cs:Baseline,16);
\coordinate (O1) at (rel axis cs:0,0);
\coordinate (O2) at (rel axis cs:1,0);
\addlegendentry{System}
    \end{axis}
\end{tikzpicture}
    \vspace{-12pt}\caption{Performance of each optimization implemented with 1 \gls{cu} and $p=11$.}\vspace{-6pt}
    \label{fig:perf_cu1p11}
\end{figure}
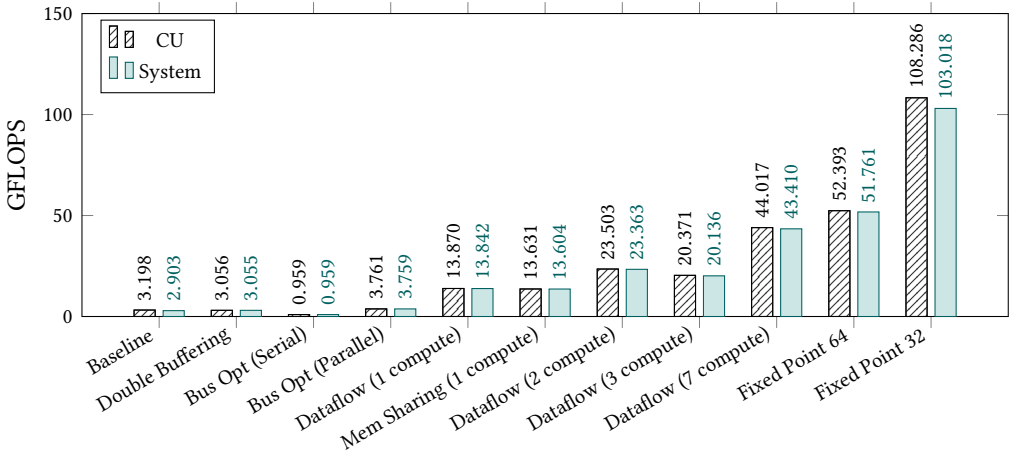

We executed our CFDlang on the DSL description in \autoref{fig:helm_op:dsl} to generate the C kernel for hardware optimization. We first performed experiments to evaluate the effects of optimizations with $p=11$. In particular, we progressively added the following optimizations:
\begin{itemize}[leftmargin=1em]
    \item \textsc{Baseline}: No optimizations are used. The code serially executes kernels and data transfers, while each \gls{cu} contains only one kernel and is connected to the \gls{hbm} with 64-bit AXI channels.
    \item \textsc{Host-HBM Double Buffering}: We introduce this well-known optimization to hide CPU-FPGA communication latency.
    \item \textsc{HBM-FPGA Bus Optimization}: We evaluate the effect of widening the bus to 256 bits, with only one kernel unit (and serializing the data) and with multiple lanes feeding parallel kernel units.
    \item \textsc{Dataflow optimization}: We create several variants of the compute functions with one, two, three, and seven subkernels. We evaluate the performance vs. resources trade-off.
    \item \textsc{Resource Optimization}: We apply on-chip memory sharing (only in the case of dataflow implementations with one block inside the compute part) and fixed-point optimizations (with 64- and 32-bit implementations). 
\end{itemize}
For each of these implementations, we measured total and kernel execution times, maximum and average power consumption, and cost in terms of hardware resources. \autoref{fig:perf_cu1p11} shows the performance (in terms of GFLOPS) achieved in each experiment when adding the specific optimization on top of the previous ones. In each experiment, the left black~\&~white bar (CU) shows the GFLOPS of the \glspl{cu} on their own, without considering host-FPGA data transfers, while the right azure bar (System) includes the entire application. Comparing the two bars allows us to evaluate the peak performance of the kernels and the effects of data transfers.

The \textsc{Baseline} case achieves only 2.9 GFLOPS, and the difference between the \gls{cu} performance and the overall system performance is 9.2\%. This is due to the serial nature of the implementation where data is transferred from the host to the \gls{hbm}, then processed by the \gls{cu} and sent back to the host before starting a new batch. If more data needs to be transferred, this discrepancy between \gls{cu} performance and overall system performance will grow larger, as the \gls{cu} needs to wait for all of the data to be sent before beginning execution.

After the \textsc{Double Buffering} optimization, the \gls{cu} performance remains similar, with a small degradation due to overhead, while the system performance is now the same as the \gls{cu} performance. This is an improvement over the \textsc{Baseline} implementation because now the host to \gls{hbm} data transfers are happening in parallel to and are entirely hidden behind the \gls{cu} execution. 

We then executed two experiments for evaluating \textsc{Bus Optimization}. In the \textit{Serial} version, we attempt to utilize the full bandwidth of the 256-bit bus by packing four doubles. The \gls{cu} reads them in parallel, but then it serializes them when it needs to access its own local buffers. While this optimization is supposed to speed up data reads from the \gls{hbm}, its implementation in the \gls{cu} leads to a performance \textit{degradation} of about 3$\times$. This is mostly due to the complexity of aligning the data ($p\times p$ and $p\times p\times p$) to multiples of fours inside the bus. To combat this, but still use the full bus bandwidth, this optimization was replaced with the \textit{Parallel} implementation where four kernels are instantiated in the \gls{cu} and the data for each \textquote{lane} is stored in separate buffers, one for each kernel. This led to a $3.92\times$ speedup over the \textit{Serial} implementation, as the four lanes are now all read in parallel. There is only a $1.23\times$ speedup over the \textsc{Double Buffering} implementation, where a $4\times$ speedup would be expected. This is because in both \textsc{Bus Opt} implementations, the \gls{hls} tool only instantiated two double-precision multipliers (rather than 11 in all other implementations). So when the innermost loops are unrolled, a resource limitation violation increases the \gls{ii} to 4, effectively counteracting the expected $4\times$ speedup.
We use the \textit{Parallel} architecture in the following experiments.

Next, we tested various forms of the \textsc{Dataflow Optimization}. Each implementation of this optimization separated the kernels into read, compute, and write modules and streams were used to pass data between them, allowing a pipelined structure. When using one compute subkernel (\textit{Dataflow (1 Compute)}) test, the speedup was $3.68\times$. In these cases, the \gls{hls} tool instantiated the full 11 multipliers, removing the \gls{ii} violation.
This decision of the HLS tool along with the overlapping execution of the read, compute, and write modules allows us to effectively achieve a 4$\times$ speed-up, i.e., to full exploit the four parallel lanes. Since the compute module dominated the execution time, it was further split into 2, 3, and 7 modules. 
The Inverse Helmholtz operator comprises seven loop nests, each implementing the operations in the seven grey rows on the right side of \autoref{fig:factorization}. To split into 2 modules, the kernel is divided into a first module with the first three loop nests with \texttt{S} and \texttt{u} as input and \texttt{t} as output, and a second module with the last four loop nests with \texttt{S}, \texttt{D}, and \texttt{t} as input and \texttt{v} as the final output. The split was made as such so that the first module does not need \texttt{D} as an input and the second module does not need \texttt{u} as an input.
To split into 3 modules, we use the division shown on the left side of \autoref{fig:factorization} and in \autoref{fig:operator_scheduling}. This is the most natural division as it matches the initial \gls{dsl} representation where the first three loop nests implement the \texttt{gemm} operator, the fourth loop nest implements the \texttt{mmult} operator, and the last three loop nests implement the \texttt{gemm\_inv} operator. Another benefit to this division is that the \texttt{mmult} loop nest consumes and produces data in the same order it is sent via the streams, meaning that no extra buffering is needed for this module and each data element can be immediately processed as it is received leading to a minimal latency. To split into 7 modules, each loop nest is a separate module.

All three of these tests gained speedup over the \textit{1-Compute} version by breaking the total execution time of a single module down further. 
The \textit{2-Compute} version is $1.7\times$ faster than the \textit{1-Compute} version. The discrepancy between this result and the ideal speedup of $2\times$ is due to the extra data buffering that must be done in each module to allow random access. In the \textit{1-Compute} case, the input arrays are each buffered one time, which adds an overall latency equal to the total input data size. In the \textit{2-Compute} case, the \texttt{S} array is needed by both modules and must be buffered twice. Additionally, the output of the first module, \texttt{t}, is used as input to the second module and must also be buffered. This extra buffering is overlapped while the two modules execute in a pipeline fashion, but it means that the latencies of each module are not exactly half of the total latency of one unified module.
However, \textit{3-Compute} modules was slower than \textit{2-Compute} modules. 
The overall execution time is determined by the module with the longest latency, as it is the limiting factor in the overall latency of the system. Because the loop nest implementing the \texttt{gemm} operator has a minimal latency, moving it to a separate module does not significantly reduce the latency of the largest module.
In fact, in each case, the module with the longest latency was the same, but the extra modules and control routing caused the tools to frequency scale the \textit{3-Compute} case to execute at 266~MHz whereas the \textit{2-Compute} case executed at 292~MHz. 
When this is considered, the performance of both tests is approximately the same. The \textit{7-Compute} test, however, performed the best because each of the compute modules were much smaller than the previous tests. In this case, the latencies of these modules were now slightly shorter than the latency of the read module, meaning that this is the limit of the performance increase by dividing the compute portion. The \textit{7-Compute} test gained a total speedup of $4.03\times$ over the \textsc{Bus Opt} \textit{Parallel} implementation. 

To evaluate the efficiency of the allocated resources, we computed the ``ideal'' GFLOPS value for each of the double-precision floating-point implementations. We identified the total number of double-precision adders and multipliers instantiated in the \gls{cu} (\# Ops) by analyzing the synthesis reports generated by Vitis HLS. The ideal GFLOPS is computed by multiplying this value by the frequency of the \gls{cu} and represents the performance when all operators were constantly in use concurrently.
\autoref{tab:metric} compares these values with the measured GFLOPS of each implementation. In the last column, an ``efficiency'' is calculated as a ratio between the ideal and achieved GFLOPS. 
\begin{table}[!ht]
    \centering
    \caption{Efficiency of floating-point operators}
    \label{tab:metric}
    \vspace{-6 pt}\rowcolors{2}{white}{gray!25}
    \resizebox{0.7\columnwidth}{!}{
    \begin{tabular}{crrrrr}
    \toprule
    %     &       & $f$   & Expected GFLOPS & Achieved &  \\ 
    & \# Ops 
    & \begin{tabular}{@{}c@{}}$f$\\(MHz)\end{tabular} 
    & \begin{tabular}{@{}c@{}}Ideal GFLOPS\\(\# Ops $\times f$)\end{tabular} 
    & \begin{tabular}{@{}c@{}}Achieved\\GFLOPS\end{tabular} 
    & Efficiency\\ 
    \midrule
        Baseline & 22 & 274.6 & 6.041 & 2.903                &0.481 \\ 
        Double Buffering & 22 & 259.8 & 5.716 & 3.055        &0.535 \\ 
        Bus Opt (Serial) & 4 & 286.5 & 1.146 & 0.959         &0.837 \\ 
        Bus Opt (Parallel) & 16 & 296.6 & 4.746 & 3.759      &0.792 \\ 
        Dataflow (1 compute) & 88 & 286.2 & 25.186 & 13.842  &0.550 \\ 
        Dataflow (2 compute) & 176 & 291.9 & 51.374 & 23.363 &0.455 \\ 
        Dataflow (3 compute) & 180 & 266.3 & 47.934 & 20.136 &0.420 \\
        Dataflow (7 compute) & 532 & 199.5 & 106.134 & 43.410&0.409 \\
        \bottomrule
    \end{tabular}
    }
\end{table}

This ``efficiency'' reflects the behavior of the allocation and scheduling of the HLS tool more than the efficiency of our system design surrounding each HLS kernel. However, we can still gain some insight into our design in the cases where the HLS decisions were the same. For instance, in the \emph{Baseline} and \emph{Double Buffering} cases, the same kernels are used, and therefore they each have the same \# Ops. The efficiency increases because less time is ``wasted'' waiting on data transfers from the host. Both \emph{Bus Opt} implementations reduce the \# Ops because the HLS tool used a different local memory type with fewer read ports, restricting the unrolling, and therefore only used two adders and two multipliers for each kernel. 
The efficiency values of these implementations are also much higher because these are the only cases where the multipliers themselves are pipelined. 
The ideal GFLOPS metric expects each operator to produce a result every clock cycle, so in all other cases where the operators are not pipelined, there are several cycles of latency for each operation reducing the efficiency.
Between each of the \emph{Dataflow} implementations, the efficiency drops slightly as the computation is split into more modules because it is impossible to split the computation into equal latency modules. The module with the longest latency may constantly be computing, but the shorter-length modules must stall. 

The efficiency values for all implementations (except \emph{Bus Opt}) are all near 0.5 because each multiply-accumulate is implemented as eleven parallel multipliers and eleven sequential adders. Even though the additions are sequential, the tool still allocated eleven of them. Because the \emph{Bus Opt} implementations are restricted to two adders, their efficiencies are higher.

\begin{table}[t]
    \centering
    \caption{Resource utilization for each optimization implemented with 1 \gls{cu} and $p=11$. Highlighted in red is any value over 25\% utilization, indicating possible issues when instantiating more than one \gls{cu}. }
    \label{tab:area}
    \vspace{-6pt}
    \resizebox{\columnwidth}{!}{
    \pgfplotstabletypeset[
        skip rows between index={9}{10},
        skip rows between index={11}{12},
        skip rows between index={13}{1000},%string type,
        every head row/.style={
            output empty row,
            before row={
                \toprule &  
                \begin{tabular}{@{}c@{}}$f_{max}$ \\ (MHz)\end{tabular}
                & \multicolumn{2}{c}{LUT} & \multicolumn{2}{c}{FF} 
                & \multicolumn{2}{c}{BRAM} 
                & \multicolumn{2}{c}{URAM} 
                & \multicolumn{2}{c}{DSP}\\
            },
            after row=\cmidrule(r){2-2}\cmidrule{3-4}\cmidrule(l){5-6}\cmidrule(l){7-8}\cmidrule(l){9-10}\cmidrule(l){11-12},
        },
        columns={test,real_freq,t_lut,t_lut_per,t_ff,t_ff_per,
            t_bram,t_bram_per, t_uram, t_uram_per,
            t_dsp,t_dsp_per},
        columns/test/.style={string type, column name={}},
        columns/real_freq/.style={fixed,column type={c@{\hspace{1em}}}},
        columns/t_lut/.style={fixed,column type={r@{\hspace{1em}}}},
        columns/t_lut/.append style={
                postproc cell content/.code={%
                \pgfkeysalso{@cell content=\pgfmathtruncatemacro\number{##1}\ifnum\number>325680\color{red}\fi##1}%
                },
        },
        columns/t_lut_per/.append style={fixed,zerofill,precision=1,column type={r},
                postproc cell content/.append code={%
                \pgfkeysgetvalue{/pgfplots/table/@cell content}{\myTmpVal}%
                \pgfkeysalso{@cell content=\pgfmathtruncatemacro\number{##1}\ifnum\number>25\color{red}\fi(##1\%)}%
                }, 
        },
     columns/t_ff/.style={fixed,column type={r@{\hspace{1em}}}},
        columns/t_ff/.append style={
                postproc cell content/.code={%
                \pgfkeysalso{@cell content=\pgfmathtruncatemacro\number{##1}\ifnum\number>651840\color{red}\fi##1}%
                },
        },
        columns/t_ff_per/.style={fixed,zerofill,precision=1,column type={r},%dec sep align,
                postproc cell content/.append code={%
                \pgfkeysalso{@cell content=\pgfmathtruncatemacro\number{##1}\ifnum\number>24\color{red}\fi(##1\%)}%
                }, 
        },
        columns/t_bram/.style={fixed,column type={r@{\hspace{1em}}}},
        columns/t_bram/.append style={
                postproc cell content/.code={%
                \pgfkeysalso{@cell content=\pgfmathtruncatemacro\number{##1}\ifnum\number>504\color{red}\fi##1}%
                },
        },
        columns/t_bram_per/.style={fixed,zerofill,precision=1,column type={r},%dec sep align,
                postproc cell content/.append code={%
                \pgfkeysalso{@cell content=\pgfmathtruncatemacro\number{##1}\ifnum\number>24\color{red}\fi(##1\%)}%
                }, 
        },
        columns/t_uram/.style={fixed,column type={r@{\hspace{1em}}}},
        columns/t_uram/.append style={
                postproc cell content/.code={%
                \pgfkeysalso{@cell content=\pgfmathtruncatemacro\number{##1}\ifnum\number>240\color{red}\fi##1}%
                },
        },
        columns/t_uram_per/.style={fixed,zerofill,precision=1,column type={r},%dec sep align,
                postproc cell content/.append code={%
                \pgfkeysalso{@cell content=\pgfmathtruncatemacro\number{##1}\ifnum\number>25\color{red}\fi(##1\%)}%
                }, 
        },
        columns/t_dsp/.style={fixed,column type={r@{\hspace{1em}}}},
        columns/t_dsp/.append style={
                postproc cell content/.code={%
                \pgfkeysalso{@cell content=\pgfmathtruncatemacro\number{##1}\ifnum\number>2256\color{red}\fi##1}%
                },
        },
        columns/t_dsp_per/.style={fixed,zerofill,precision=1,column type={r},%dec sep align,
                postproc cell content/.append code={%
                \pgfkeysalso{@cell content=\pgfmathtruncatemacro\number{##1}\ifnum\number>24\color{red}\fi(##1\%)}%
                }, 
        },
    ]{\pelevencuonetable}}\vspace{-6pt}
\end{table}

At this point, we want to start replicating the \glspl{cu} using the remaining area available in the FPGA fabric to maximize parallelism. For this reason, we need to evaluate the hardware cost of each implementation. The numbers of LUT, FF, BRAM, URAM, and DSP used by each case for $p=11$ are shown in \autoref{tab:area}.
In general, each test from Baseline to Dataflow (7 Compute) showed an increase in resource utilization. Any utilization value over 25\% is shown in red. These resources are most likely to cause placement and routing issues when instantiating multiple \glspl{cu}. We tested a few methods to reduce resource utilization to be able to increase the number of instantiated \glspl{cu}.

The \textit{Mem Sharing} optimization is applied to the \textsc{Dataflow} \textit{1-Compute} implementation where several arrays are used in the compute module (cf. \autoref{fig:sharing}). Mnemosyne generated an architecture to internally share arrays based on their liveness intervals.
This decreased the BRAM utilization by 14.5\% and the URAM utilization by 48.3\% while the LUT and FF utilization only increased minimally and the DSP utilization remained the same. Also, the execution time was only slightly reduced (a slowdown of $0.98\times$). Conversely, this optimization cannot be applied to the \textsc{Dataflow} \textit{2-Compute}, \textit{3-Compute}, and \textit{7-Compute} implementations because, in these cases, each compute module only uses arrays that cannot be shared, as they are always in use during the module execution. This optimization is indeed beneficial when on-chip memory inside the \gls{cu} is the limiting factor, and when replicating the \glspl{cu} brings more improvements than dataflow execution.

Another method to reduce resources is to change the numerical representation. All of the previous tests used the floating-point format with double precision. In general, fixed-point representations utilize fewer resources than floating-point ones. We tested 64- and 32-bit fixed-point representations by modifying the \textsc{Dataflow} \textit{7-Compute} implementation. The 64-bit implementation uses 24 bits for the integer portion and 40 bits for the fractional portion. The 32-bit implementation uses 8 bits for the integer portion and 24 bits for the fractional portion. These values are provided by the user after an analysis of the algorithm.
Because the 32-bit data is half the size, we instantiate eight kernels per \gls{cu} and divide the 256-bit bus into eight lanes. 
In the \textit{Fixed Point 64} test, the LUT utilization reduced by 46.3\%, the FF utilization reduced by 53.4\%, the RAM utilization remained the same, and the DSP utilization increased by 44.8\%. 
In the \textit{Fixed Point 32} test, concerning the \textit{Fixed Point 64} test, the LUT and FF utilization remained roughly the same. The DSP utilization was nearly halved. The BRAM increased by about four times while the URAM decreased to zero. This is because the data representation is half as long, so the overall size of the data structures is half as big. The arrays representing the tensors are no longer big enough for the synthesis tool to decide if it is efficient to use URAM to store them. When considering the size of the physical memories, the total memory space is approximately halved.
The performance of the \textit{Fixed Point 64} test had a slight speedup of $1.19\times$ due to the simplification of the logic allowing the frequency to be higher. The \textsc{Dataflow} \textit{7-Compute} test with double format was scaled to 199~MHz while the \textit{Fixed Point 64} test was scaled to 234~MHz. The performance of the \textit{Fixed Point 32} test had a speedup over the double format of $2.37\times$ and it reaches up to 103~GFLOPS. This represents a speed up of more than 35$\times$ over the \textsc{Baseline} version. The \textit{Fixed Point 64} test exhibited a mean square error of $9.39\times10^{-22}$ while the \textit{Fixed Point 32} test had a mean square error of $3.58\times10^{-12}$. It is up to the application designer to determine what an acceptable error is and decide on an appropriate number format, and our flow can help facilitate a design space exploration of these parameters.

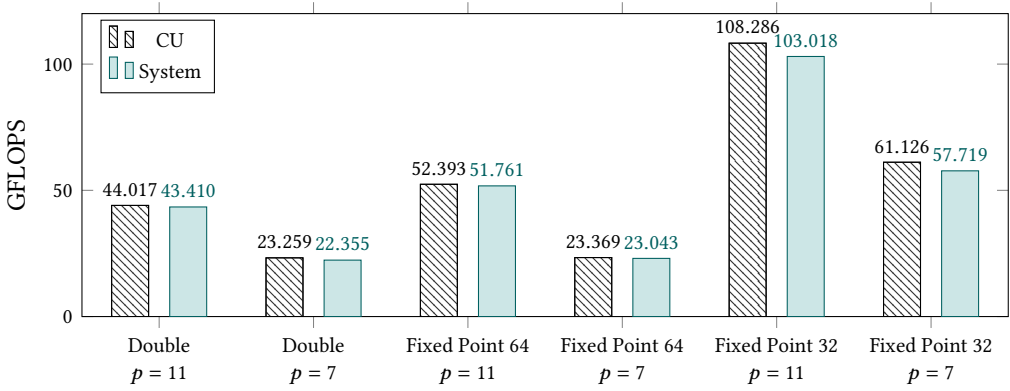
\begin{figure}[t]
    \centering
\begin{tikzpicture}
    \begin{axis}[
    symbolic x coords=
        {
            Baseline,
            Double Buffering,
            Bus Opt (Serial),
            Bus Opt (Parallel),
            Dataflow (1 compute),
            Dataflow (2 compute),
            Dataflow (3 compute),
            Dataflow (7 compute),
            Double p7,
            Fixed Point 64,
            Fixed Point 64 p7,
            Fixed Point 32,
            Fixed Point 32 p7,
            Mem Sharing (1 compute),
            Dataflow (7 compute) - Far,
            Double p11,
            Fixed Point 64 p11,
            Fixed Point 32 p11,           
            Double p11cu2,
            Fixed Point 64 p11cu2,
            Fixed Point 32 p11cu3,
            Double p7cu3,
            Fixed Point 64 p7cu2,
            Fixed Point 32 p7cu4,
        },
            ylabel={GFLOPS},
            ymin=0,
            ymax=120,
            x tick label style={
            font=\footnotesize,
            },
            every tick label/.append style={font=\footnotesize},
            xtick=data,
            xticklabel style   = {align=center},
            xticklabels={
                Double\\$p=11$,
                Double\\$p=7$,
                Fixed Point 64\\$p=11$,
                Fixed Point 64\\$p=7$,
                Fixed Point 32\\$p=11$,
                Fixed Point 32\\$p=7$
            },
            legend style={ font=\footnotesize,at={(0.02,0.98)},anchor=north west},
            width=1\columnwidth, height=0.4\textwidth,
            ybar,
            bar width=14pt,
            nodes near coords,
            nodes near coords style={
            font=\footnotesize,
            /pgf/number format/fixed, /pgf/number format/precision=3, /pgf/number format/zerofill},
            bar shift auto=8pt,
        ]
        \addplot[black,
        x filter/.code={
        \ifnum\coordindex<7\def\pgfmathresult{}\fi
        \ifnum\coordindex=8\def\pgfmathresult{}\fi
        \ifnum\coordindex>13\def\pgfmathresult{}\fi
        },
        postaction={pattern=north west lines,pattern color=black!80}] table[x=test, y={k_gflops}] {\pelevencuonetable};
        \addlegendentry{CU}
        \addplot[black!25!teal,
        x filter/.code={
        \ifnum\coordindex<7\def\pgfmathresult{}\fi
        \ifnum\coordindex=8\def\pgfmathresult{}\fi
        \ifnum\coordindex>13\def\pgfmathresult{}\fi
        },
        fill=teal!20] table[x=test, y={sys_gflops}] {\pelevencuonetable};
        \addlegendentry{System}
    \end{axis}

\end{tikzpicture}
    \vspace{-12pt}\caption{Performance of each data representation implemented with 1 \gls{cu} and $p=11$ or $p=7$}
    \label{fig:perf_cu1p7}
\end{figure}

Another method to reduce resource utilization for this kernel is to vary the input parameter $p$. We tested the \textsc{Dataflow} \textit{7-Compute} implementation using 64-bit double, 64-bit fixed point, and 32-bit fixed point with $p=7$ and $p=11$. The results are summarized in \autoref{fig:perf_cu1p7} and \autoref{tab:areap7}. 

\begin{table}[t]
    \centering
    \caption{Resource utilization for each data representation implemented with 1 \gls{cu} and $p=11$ or $p=7$. Shown in red is any value over 25\% utilization, indicating possible issues when instantiating multiple \glspl{cu}.}
    \label{tab:areap7}\vspace{-5pt}
    \resizebox{\columnwidth}{!}{
    \pgfplotstabletypeset[%string type,
        skip rows between index={0}{7},
                skip rows between index={8}{9},
        skip rows between index={14}{1000},
        every head row/.style={
            output empty row,
            before row={
                \toprule & $p$ &
                \begin{tabular}{@{}c@{}}$f_{max}$ \\ (MHz)\end{tabular}
                & \multicolumn{2}{c}{LUT} & \multicolumn{2}{c}{FF}
                & \multicolumn{2}{c}{BRAM}
                & \multicolumn{2}{c}{URAM}
                & \multicolumn{2}{c}{DSP}\\
            },
            after row=\cmidrule(r){2-2}\cmidrule(r){3-3}\cmidrule{4-5}\cmidrule(l){6-7}\cmidrule(l){8-9}\cmidrule(l){10-11}\cmidrule(l){12-13},
        },
        columns={newcol,p,real_freq,t_lut,t_lut_per,t_ff,t_ff_per,
            t_bram,t_bram_per, t_uram, t_uram_per,
            t_dsp,t_dsp_per},
        create on use/newcol/.style={
            create col/set list={0,1,2,3,4,5,6,Double,Double,Double,Fixed Point 64,Fixed Point 64,Fixed Point 32, Fixed Point 32}
        },
        columns/newcol/.style={string type, column name={}},
        columns/p/.style={fixed,column type={c@{\hspace{1em}}}},
        columns/real_freq/.style={fixed,zerofill,precision=1,column type={c@{\hspace{1em}}}},
        %columns/test/.style={string type, column name={}},
                columns/t_lut/.style={fixed,column type={r@{\hspace{1em}}}},
        columns/t_lut/.append style={
                postproc cell content/.code={%
                \pgfkeysalso{@cell content=\pgfmathtruncatemacro\number{##1}\ifnum\number>325680\color{red}\fi##1}%
                },
        },
        columns/t_lut_per/.append style={fixed,zerofill,precision=1,column type={r},
                postproc cell content/.append code={%
                \pgfkeysgetvalue{/pgfplots/table/@cell content}{\myTmpVal}%
                \pgfkeysalso{@cell content=\pgfmathtruncatemacro\number{##1}\ifnum\number>24\color{red}\fi(##1\%)}%
                }, %NOTE this doesn't actually set the precision
        },
     columns/t_ff/.style={fixed,column type={r@{\hspace{1em}}}},
        columns/t_ff/.append style={
                postproc cell content/.code={%
                \pgfkeysalso{@cell content=\pgfmathtruncatemacro\number{##1}\ifnum\number>651840\color{red}\fi##1}%
                },
        },
        columns/t_ff_per/.style={fixed,zerofill,precision=1,column type={r},%dec sep align,
                postproc cell content/.append code={%
                \pgfkeysalso{@cell content=\pgfmathtruncatemacro\number{##1}\ifnum\number>24\color{red}\fi(##1\%)}%
                }, %NOTE this doesn't actually set the precision
        },
        columns/t_bram/.style={fixed,column type={r@{\hspace{1em}}}},
        columns/t_bram/.append style={
                postproc cell content/.code={%
                \pgfkeysalso{@cell content=\pgfmathtruncatemacro\number{##1}\ifnum\number>504\color{red}\fi##1}%
                },
        },
        columns/t_bram_per/.style={fixed,zerofill,precision=1,column type={r},%dec sep align,
                postproc cell content/.append code={%
                \pgfkeysalso{@cell content=\pgfmathtruncatemacro\number{##1}\ifnum\number>24\color{red}\fi(##1\%)}%
                }, %NOTE this doesn't actually set the precision
        },
        columns/t_uram/.style={fixed,column type={r@{\hspace{1em}}}},
        columns/t_uram/.append style={
                postproc cell content/.code={%
                \pgfkeysalso{@cell content=\pgfmathtruncatemacro\number{##1}\ifnum\number>240\color{red}\fi##1}%
                },
        },
        columns/t_uram_per/.style={fixed,zerofill,precision=1,column type={r},%dec sep align,
                postproc cell content/.append code={%
                \pgfkeysalso{@cell content=\pgfmathtruncatemacro\number{##1}\ifnum\number>25\color{red}\fi(##1\%)}%
                }, %NOTE this doesn't actually set the precision
        },
        columns/t_dsp/.style={fixed,column type={r@{\hspace{1em}}}},
        columns/t_dsp/.append style={
                postproc cell content/.code={%
                \pgfkeysalso{@cell content=\pgfmathtruncatemacro\number{##1}\ifnum\number>2256\color{red}\fi##1}%
                },
        },
        columns/t_dsp_per/.style={fixed,zerofill,precision=1,column type={r},%dec sep align,
                postproc cell content/.append code={%
                \pgfkeysalso{@cell content=\pgfmathtruncatemacro\number{##1}\ifnum\number>24\color{red}\fi(##1\%)}%
                }, %NOTE this doesn't actually set the precision
        },
    ]{\pelevencuonetable}}
    \vspace{-6pt}
\end{table}

Compared to their $p=11$ counterparts, the $p=7$ implementations performed slightly slower. 
This is because the actual hardware implementation does not scale the same way as the conceptual floating point operations per kernel (used to compute the GFLOPS). However, the resource reduction between $p=11$ and $p=7$ is enough to allow for more replication of the \glspl{cu}. For instance, the \textit{Fixed Point 32} implementation uses 66.4\% of the available BRAM for $p=11$ while it only uses 21.7\% for $p=7$, allowing $4\times$ replication.

To further facilitate instantiating multiple \glspl{cu}, we reduced the stream FIFOs from a naive full size to a small enough version to save space and still prevent deadlock. This led to a small performance reduction due to stalls but significantly reduced the total number of BRAMs. Also, because the DSP utilization was exceptionally high in some cases, we used pragmas to guide the \gls{hls} tool on using LUTs instead of DSPs to implement fixed-point multipliers. We used this pragma in one of the seven compute modules to shift some of the resource load off of DSPs and onto LUTs. 

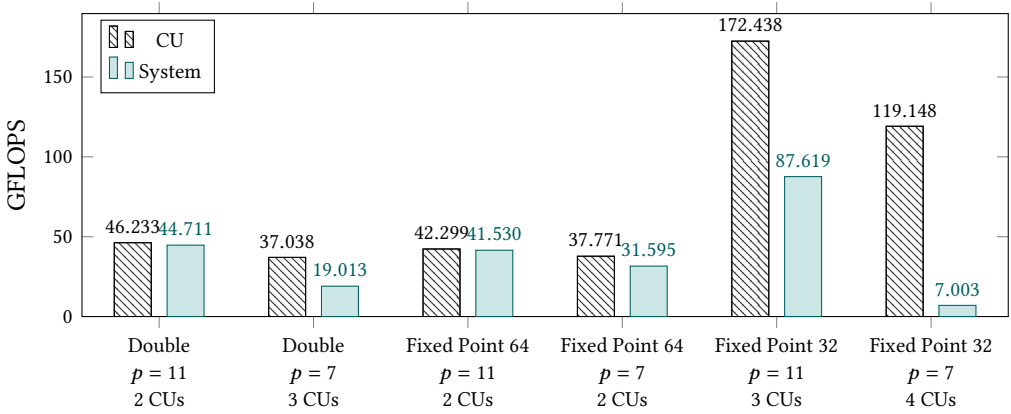
\begin{figure}[t]
    \centering
\begin{tikzpicture}
    \begin{axis}[
    symbolic x coords=
        {
            Baseline,
            Double Buffering,
            Bus Opt (Serial),
            Bus Opt (Parallel),
            Dataflow (1 compute),
            Dataflow (2 compute),
            Dataflow (3 compute),
            Dataflow (7 compute),
            Double p7,
            Fixed Point 64,
            Fixed Point 64 p7,
            Fixed Point 32,
            Fixed Point 32 p7,
            Mem Sharing (1 compute),
            Dataflow (7 compute) - Far,
            Double p11,
            Fixed Point 64 p11,
            Fixed Point 32 p11,           
            Double p11cu2,
            Double p7cu3,
            Fixed Point 64 p11cu2,
            Fixed Point 64 p7cu2,
            Fixed Point 32 p11cu3,
            Fixed Point 32 p7cu4,
        },
            ylabel={GFLOPS},
            ymin=0,
            x tick label style={
            font=\footnotesize
            },
            every tick label/.append style={font=\footnotesize},
            xtick=data,
            xticklabel style   = {align=center},
            xticklabels={
                Double\\$p=11$\\2 CUs,
                Double\\$p=7$\\3 CUs,
                Fixed Point 64\\$p=11$\\2 CUs,
                Fixed Point 64\\$p=7$\\2 CUs,
                Fixed Point 32\\$p=11$\\3 CUs,
                Fixed Point 32\\$p=7$\\4 CUs
            },
            legend style={ font=\footnotesize,at={(0.02,0.98)},anchor=north west},
            width=1\columnwidth, height=0.4\textwidth,
            ybar,
            bar width=14pt,
            nodes near coords,
            nodes near coords style={
            font=\footnotesize,
            /pgf/number format/fixed, /pgf/number format/precision=3, /pgf/number format/zerofill},
            bar shift auto=6pt,
        ]
        \addplot[black,
        x filter/.code={
        \ifnum\coordindex<14\def\pgfmathresult{}\fi
        \ifnum\coordindex>19\def\pgfmathresult{}\fi
        },
        postaction={pattern=north west lines,pattern color=black!80}] table[x=test, y={k_gflops}] {\pelevencuonetable};
        \addlegendentry{CU}
        \addplot[black!25!teal,
        x filter/.code={
        \ifnum\coordindex<14\def\pgfmathresult{}\fi
        \ifnum\coordindex>19\def\pgfmathresult{}\fi
        },
        fill=teal!20] table[x=test, y={sys_gflops}] {\pelevencuonetable};
        \addlegendentry{System}
    \end{axis}

\end{tikzpicture}
    \vspace{-6pt}\caption{Performance of each data representation implemented with multiple \glspl{cu} and $p=11$ or $p=7$}
    \label{fig:perf_cumultp7}
\end{figure}

We were able to instantiate two parallel \glspl{cu} for the cases of \textit{Double} with $p=11$, \textit{Fixed Point 64} with $p=11$, and \textit{Fixed Point 64} with $p=7$, three \glspl{cu} for the cases of \textit{Double} with $p=7$ and \textit{Fixed Point 32} with $p=11$, and four \glspl{cu} for the case of \textit{Fixed Point 32} with $p=7$. The performance results for these implementations are shown in \autoref{fig:perf_cumultp7}, and the area results are shown in \autoref{tab:areacumultp7}. All of these implementations were built targeting 225~MHz, as most of their 1~\gls{cu} counterparts could not even achieve this. 

\begin{table}[t]
    \centering
    \caption{Resource utilization for each data representation implemented with multiple \glspl{cu} and $p=11$ or $p=7$. Shown in red is any value over 50\% utilization to show which resources are the limiting factor.}
    \label{tab:areacumultp7}\vspace{-5pt}
    \resizebox{\columnwidth}{!}{
    \pgfplotstabletypeset[%string type,
        skip rows between index={0}{14},
        skip rows between index={20}{1000},
        every head row/.style={
            output empty row,
            before row={
                \toprule & $p$ & CUs&
                \begin{tabular}{@{}c@{}}$f_{max}$ \\ (MHz)\end{tabular}
                & \multicolumn{2}{c}{LUT} & \multicolumn{2}{c}{FF}
                & \multicolumn{2}{c}{BRAM}
                & \multicolumn{2}{c}{URAM}
                & \multicolumn{2}{c}{DSP}\\
            },
            after row=\cmidrule(r){2-2}\cmidrule(r){3-3}\cmidrule(r){4-4}\cmidrule{5-6}\cmidrule(l){7-8}\cmidrule(l){9-10}\cmidrule(l){11-12}\cmidrule(l){13-14},
        },
        columns={newcol,p,cu,real_freq,t_lut,t_lut_per,t_ff,t_ff_per,
            t_bram,t_bram_per, t_uram, t_uram_per,
            t_dsp,t_dsp_per},
        create on use/newcol/.style={
            create col/set list={0,1,2,3,4,5,6,Double,Double,Double,Fixed Point 64,Fixed Point 64,Fixed Point 32, Fixed Point 32,Double,Double,Fixed Point 64,Fixed Point 64,Fixed Point 32, Fixed Point 32}
        },
        columns/newcol/.style={string type, column name={}},
        columns/p/.style={fixed,column type={c@{\hspace{1em}}}},
        columns/cu/.style={fixed,column type={c@{\hspace{1em}}}},
        columns/real_freq/.style={fixed,zerofill,precision=1,column type={c@{\hspace{1em}}}},
                columns/t_lut/.style={fixed,column type={r@{\hspace{1em}}}},
        columns/t_lut/.append style={
                postproc cell content/.code={%
                \pgfkeysalso{@cell content=\pgfmathtruncatemacro\number{##1}\ifnum\number>651360\color{red}\fi##1}%
                },
        },
        columns/t_lut_per/.append style={fixed,zerofill,precision=1,column type={r},
                postproc cell content/.append code={%
                \pgfkeysalso{@cell content=\pgfmathtruncatemacro\number{##1}\ifnum\number>49\color{red}\fi(##1\%)}%
                %\pgfkeysalso{@cell content=(##1\%)}%
                }, %NOTE this doesn't actually set the precision
        },
     columns/t_ff/.style={fixed,column type={r@{\hspace{1em}}}},
        columns/t_ff/.append style={
                postproc cell content/.code={%
                \pgfkeysalso{@cell content=\pgfmathtruncatemacro\number{##1}\ifnum\number>1303680\color{red}\fi##1}%
                },
        },
        columns/t_ff_per/.style={fixed,zerofill,precision=1,column type={r},%dec sep align,
                postproc cell content/.append code={%
                \pgfkeysalso{@cell content=\pgfmathtruncatemacro\number{##1}\ifnum\number>49\color{red}\fi(##1\%)}%
                %\pgfkeysalso{@cell content=(##1\%)}%
                }, %NOTE this doesn't actually set the precision
        },
        columns/t_bram/.style={fixed,column type={r@{\hspace{1em}}}},
        columns/t_bram/.append style={
                postproc cell content/.code={%
                \pgfkeysalso{@cell content=\pgfmathtruncatemacro\number{##1}\ifnum\number>1008\color{red}\fi##1}%
                },
        },
        columns/t_bram_per/.style={fixed,zerofill,precision=1,column type={r},%dec sep align,
                postproc cell content/.append code={%
                \pgfkeysalso{@cell content=\pgfmathtruncatemacro\number{##1}\ifnum\number>49\color{red}\fi(##1\%)}%
                %\pgfkeysalso{@cell content=(##1\%)}%
                }, %NOTE this doesn't actually set the precision
        },
        columns/t_uram/.style={fixed,column type={r@{\hspace{1em}}}},
        columns/t_uram/.append style={
                postproc cell content/.code={%
                \pgfkeysalso{@cell content=\pgfmathtruncatemacro\number{##1}\ifnum\number>480\color{red}\fi##1}%
                },
        },
        columns/t_uram_per/.style={fixed,zerofill,precision=1,column type={r},%dec sep align,
                postproc cell content/.append code={%
                \pgfkeysalso{@cell content=\pgfmathtruncatemacro\number{##1}\ifnum\number>49\color{red}\fi(##1\%)}%
                %\pgfkeysalso{@cell content=(##1\%)}%
                }, %NOTE this doesn't actually set the precision
        },
        columns/t_dsp/.style={fixed,column type={r@{\hspace{1em}}}},
        columns/t_dsp/.append style={
                postproc cell content/.code={%
                \pgfkeysalso{@cell content=\pgfmathtruncatemacro\number{##1}\ifnum\number>4512\color{red}\fi##1}%
                },
        },
        columns/t_dsp_per/.style={fixed,zerofill,precision=1,column type={r},%dec sep align,
                postproc cell content/.append code={%
                \pgfkeysalso{@cell content=\pgfmathtruncatemacro\number{##1}\ifnum\number>49\color{red}\fi(##1\%)}%
                %\pgfkeysalso{@cell content=(##1\%)}%
                }, %NOTE this doesn't actually set the precision
        },
    ]{\pelevencuonetable}}
\end{table}

In most cases, replicating the \glspl{cu} actually led to a slowdown. This is because the extra logic and routing caused the maximum frequency to be reduced thereby slowing down everything in the system. However, most cases did show speedup in terms of the \gls{cu} execution time. In particular, the \textit{Fixed Point 32} implementations achieved up to 172 GFLOPS for the kernel but around 87 GFLOPS for the system. This huge discrepancy is because even though several \glspl{cu} are now executing in parallel, all of the data must still be sent from the host to the \gls{hbm} in series. Host data transfers are now the dominating factor by far, so it is not recommended to replicate \glspl{cu} until the host data transfer time can be reduced. Otherwise, the overall system will have a slowdown from the extra logic. 

From the resource utilization results, it can be seen that both 64-bit data types are constrained by resources used for computation, namely LUTs and DSPs. The 32-bit fixed-point implementation is also somewhat constrained by DSPs. In any case, this application is composed of almost entirely floating or fixed point multiplications, and performance-optimized designs will quickly use most of the available DSPs. The 32-bit cases are also constrained by the on-chip memories, the 3 \gls{cu} implementation of \textit{Fixed Point 32} with $p=11$ even uses 100\% of the URAM, but both \textit{Fixed Point 32} implementations were able to be replicated more than their \textit{Fixed Point 64} counterparts, due to the data width reduction. The $p=7$ tests were also, in general, able to be replicated more than their $p=11$ counterparts due to the effect $p$ has on the amount of computations and the array sizes. \textit{Fixed Point 64} was only able to be replicated twice in both cases of $p$, as the reduction of DSPs between $p=11$ and $p=7$ was not enough to allow for a third \gls{cu}. 

\begin{figure}[t]
    \centering
\begin{tikzpicture}
    \begin{axis}[
    symbolic x coords=
        {
            Baseline,
            Double Buffering,
            Bus Opt (Serial),
            Bus Opt (Parallel),
            Dataflow (1 compute),
            Mem Sharing (1 compute),
            Dataflow (2 compute),
            Dataflow (3 compute),
            Dataflow (7 compute),
            Double p7,
            Fixed Point 64,
            Fixed Point 64 p7,
            Fixed Point 32,
            Fixed Point 32 p7,
            Double p11cu2,
            Double p7cu3,
            Fixed Point 64 p11cu2,
            Fixed Point 64 p7cu2,
            Fixed Point 32 p11cu3,
            Fixed Point 32 p7cu4,
            Dataflow (7 compute) - Far,
        },
            ylabel={Power (W)},
            axis y line*=left,
            ymin=0,
            ymax=65,
            x tick label style={rotate=30,anchor=east,font=\footnotesize},
            every tick label/.append style={font=\footnotesize},
            xtick=data,
            xticklabels={
                {Double, $p=11$, 1 CU},
                \textcolor{black!75}{Double, $p=7$, 1 CU},
                {Fixed Pt 64, $p=11$, 1 CU},
                \textcolor{black!75}{Fixed Pt 64, $p=7$, 1 CU},
                {Fixed Pt, 32 $p=11$, 1 CU},
                \textcolor{black!75}{Fixed Pt, 32 $p=7$, 1 CU},
                {Double, $p=11$, 2 CU},
                \textcolor{black!75}{Double, $p=7$, 3 CU},
                {Fixed Pt 64, $p=11$, 2 CU},
                \textcolor{black!75}{Fixed Pt 64, $p=7$, 2 CU},
                {Fixed Pt 32, $p=11$, 3 CU},
                \textcolor{black!75}{Fixed Pt 32, $p=7$, 4 CU}
            },
            legend style={ font=\footnotesize,at={(0.02,0.98)},anchor=north west},
            width=1\columnwidth, height=0.5\textwidth,
            ybar,
            bar width=8pt,
            nodes near coords,
            nodes near coords style={rotate=90,anchor=west,font=\footnotesize,/pgf/number format/fixed, /pgf/number format/precision=3, /pgf/number format/zerofill},
        ]
                \addlegendimage{empty legend}
        \addlegendentry{\hspace{-.6cm}\textbf{W}}
        \addplot[blue,x filter/.code={
        \ifnum\coordindex<7\def\pgfmathresult{}\fi
        \ifnum\coordindex=8\def\pgfmathresult{}\fi
        \ifnum\coordindex>19\def\pgfmathresult{}\fi
        },
        fill=blue!10, bar shift=-6pt] table[x=test, y={avg_pow}] {\pelevencuonetable};
        \addlegendentry{Average}
    \end{axis}
    \begin{axis}[
    symbolic x coords=
        {
            Baseline,
            Double Buffering,
            Bus Opt (Serial),
            Bus Opt (Parallel),
            Dataflow (1 compute),
            Mem Sharing (1 compute),
            Dataflow (2 compute),
            Dataflow (3 compute),
            Dataflow (7 compute),
            Double p7,
            Fixed Point 64,
            Fixed Point 64 p7,
            Fixed Point 32,
            Fixed Point 32 p7,
            Double p11cu2,
            Double p7cu3,
            Fixed Point 64 p11cu2,
            Fixed Point 64 p7cu2,
            Fixed Point 32 p11cu3,
            Fixed Point 32 p7cu4,
            Dataflow (7 compute) - Far,
        },
            ylabel={Efficiency (GFLOPS/W -- GOPS/W)},
            ymin=0,
            ymax=5,
            hide x axis,
            axis y line*=right,
            every tick label/.append style={font=\footnotesize},
            xtick=data,
            legend style={ font=\footnotesize,at={(0.98,0.98)},anchor=north east},
            width=1\columnwidth, height=0.5\textwidth,
            ybar,
            bar width=8pt,
            nodes near coords style={rotate=90,anchor=west,font=\footnotesize,/pgf/number format/fixed, /pgf/number format/precision=3, /pgf/number format/zerofill},
            bar shift auto=2pt,
        ]
        \addlegendimage{empty legend}
        \addlegendentry{\hspace{-.3cm}\textbf{GFLOPS/W -- GOPS/W}}
        \addplot[x filter/.code={
        \ifnum\coordindex<7\def\pgfmathresult{}\fi
        \ifnum\coordindex=8\def\pgfmathresult{}\fi
        \ifnum\coordindex>19\def\pgfmathresult{}\fi
        },black!50!green,
        ,nodes near coords,postaction={pattern=horizontal lines,pattern color=black!50!green}, bar shift=6pt] table[x=test, y={gflops_per_w}] {\pelevencuonetable};
        \addlegendentry{Alveo}
        \end{axis}
\end{tikzpicture}
    \vspace{-20pt}\caption{Power usage of the Dataflow (7 Compute) optimization with each datatype, $p=11$ or $p=7$, and 1-\gls{cu} or multiple-\gls{cu}.\vspace{-5pt}}
    \label{fig:pow_cu1p11}
\end{figure}
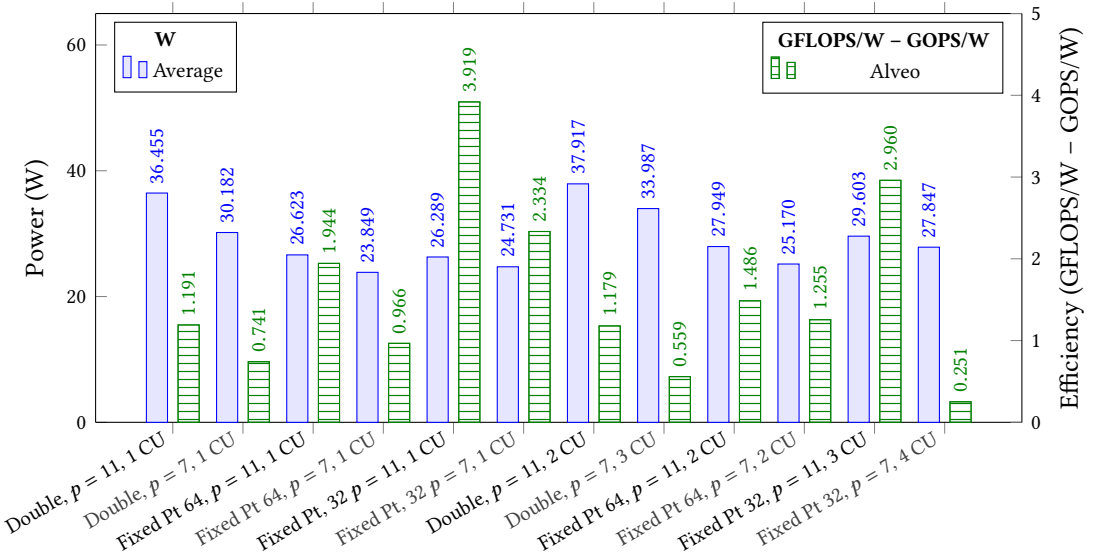

\autoref{fig:pow_cu1p11} shows the power consumption of the different implementations and a comparison of the energy efficiency (GFLOPS/W for floating-point operations and GOPS/W for fixed-point operations). 
The bars report the average power consumption measured with the XRT infrastructure. 
We also include the results of the multiple-\gls{cu} implementations to show the effects of replication on both power consumption (W bars) and energy efficiency (GFLOPS/W and GOPS/W bars). 

As expected, the fixed-point implementations are more efficient than the floating-point ones. Also, reducing the bitwidth from 64 to 32 bits allows us to achieve maximum efficiency. This is because these implementations are much faster and use fewer hardware resources. The $p=7$ implementations have lower average power consumption than their $p=11$ counterparts due to their smaller resource utilization. However, in most cases, the efficiency of the $p=7$ cases is lower due to their longer overall execution time. The multiple-\gls{cu} implementations are generally less efficient than their single-\gls{cu} counterparts, both because of the increased work occurring in parallel, yielding a higher average power, and because of longer execution times from frequency scaling. 

\subsection{Comparison with Software Implementations}\label{sec:sw_impl}
This section presents the results for two additional kernels. The first kernel performs an interpolation, which maps from $\mathbf{u}\in\mathbb{R}^{N \times N \times N}$ to $\mathbf{u}^\prime\in\mathbb{R}^{M \times M \times M}$ via an isotropic operator $\mathbf{A}\in\mathbb{R}^{M \times N}$. We implemented the \textit{Interpolation} kernel with $M=N=11$.
The second kernel computes $\nabla u$, the gradient of $u$ in all 3 dimensions. We implemented the \textit{Gradient} kernel with dimensions $8\times7\times6$. 
In all cases, we compute the total number of floating-point operations needed for simulating 2,000,000 elements, and we use these numbers to compute the GFLOPS and GFLOPS/W metrics.

\begin{figure}[t]
     \begin{subfigure}[t]{0.49\columnwidth}
    \centering
\begin{tikzpicture}
\hspace{-5pt}
    \begin{axis}[
    symbolic x coords=
        {
            Inv Helmholtz,
            Interpolation,
            Gradient,
        },
            ylabel={GFLOPS},
            ylabel shift={-4pt},
            ymin=0,
            ymax=60,
            x tick label style={rotate=0,anchor=north,font=\footnotesize},
            every tick label/.append style={font=\footnotesize},
            xtick=data,
            legend style={ font=\footnotesize,at={(0.98,0.98)},anchor=north east},
            width=1\columnwidth, height=1\columnwidth,%0.5\textwidth,
            ybar,
            bar width=8pt,
            nodes near coords,
            nodes near coords style={rotate=90, anchor=west,font=\footnotesize,/pgf/number format/fixed, /pgf/number format/precision=3, /pgf/number format/zerofill
            },
            bar shift auto=3pt,
            enlarge x limits={abs=0.85cm},
        ]
        \addplot[black,
       postaction={pattern=north east lines,pattern color=black!80}] table[x=test, y={amd_gflops}] {\addlkernelstable};
        \addlegendentry{\amd}
         \addplot[black!50!green,
        ,fill=green!20,postaction={pattern=grid,pattern color=black!50!green!80} ] table[x=test, y={naive_sys_gflops}] {\addlkernelstable};
        \addlegendentry{FPGA Baseline}
         \addplot[black!25!teal,
        ,fill=teal!20] table[x=test, y={opt_sys_gflops}] {\addlkernelstable};
        \addlegendentry{FPGA Opt.}
        \addplot[red,postaction={pattern=crosshatch,pattern color=red!80},
        ,fill=red!25,nodes near coords style={/pgf/number format/precision=0}] table[x=test, y={intel_gflops}] {\addlkernelstable};
        \addlegendentry{Intel Opt.}
    \end{axis}
\end{tikzpicture}
    \caption{Performance 
    }\vspace{-3pt}
    \label{fig:addlkernels}
     \end{subfigure}
     \hfill
     \begin{subfigure}[t]{0.49\columnwidth}
    \centering
\begin{tikzpicture}
    \begin{axis}[
    symbolic x coords=
        {
            Inv Helmholtz,
            Interpolation,
            Gradient,
        },
            ylabel={Power (W)},
            ylabel shift={-4pt},
            axis y line*=left,
            ymin=0,
            ymax=65,
            x tick label style={anchor=north,font=\footnotesize},
            every tick label/.append style={font=\footnotesize},
            xtick=data,
            legend style={ font=\footnotesize,at={(0.02,0.98)},anchor=north west},
            width=1\columnwidth, height=1\columnwidth,%0.5\textwidth,
            ybar,
            bar width=8pt,
            nodes near coords,
            nodes near coords style={rotate=90,
            anchor=west,font=\footnotesize,/pgf/number format/fixed, /pgf/number format/precision=3, /pgf/number format/zerofill
            },
            bar shift auto=3pt,
            enlarge x limits={abs=.65cm},
        ]
         \addlegendimage{empty legend}
        \addlegendentry{\hspace{-.6cm}\textbf{W}}
         \addplot[%black!25!teal,
        blue, fill=blue!10,
        bar shift=-11pt] table[x=test, y={opt_avg_pow}] {\addlkernelstable};
        \addlegendentry{FPGA Opt.}
    \end{axis}    
    \begin{axis}[
    symbolic x coords=
        {
            Inv Helmholtz,
            Interpolation,
            Gradient,
        },
            ylabel={Efficiency (GFLOPS/W)},
            ylabel shift={-4pt},
            ymin=0,
            ymax=1.8,
            x tick label style={anchor=north,font=\footnotesize},
            every tick label/.append style={font=\footnotesize},
            xtick=data,
            legend style={ font=\footnotesize,at={(0.98,0.98)},anchor=north east},
            width=1\columnwidth, height=1\columnwidth,%0.5\textwidth,
            ybar,
            bar width=8pt,
            nodes near coords,
            nodes near coords style={rotate=90,
            anchor=west,font=\footnotesize,/pgf/number format/fixed, /pgf/number format/precision=3, /pgf/number format/zerofill
            },
            bar shift auto=3pt,
            hide x axis,
            axis y line*=right,
            enlarge x limits={abs=.65cm},
        ]
        \addlegendimage{empty legend}
        \addlegendentry{\hspace{-.3cm}\textbf{GFLOPS/W}}
         \addplot[
        black!50!green ,postaction={pattern=horizontal lines,pattern color=black!50!green},
        bar shift=0pt] table[x=test, y={opt_gflops_per_w}] {\addlkernelstable};
        \addlegendentry{FPGA Opt.}
        \addplot[red,postaction={pattern=crosshatch,pattern color=red!80},
        ,fill=red!25,nodes near coords style={/pgf/number format/precision=2},bar shift=11pt] table[x=test, y={intel_gflops_per_w}] {\addlkernelstable};
        \addlegendentry{Intel Opt.}
    \end{axis}
\end{tikzpicture}
    \caption{Power usage and efficiency}\vspace{-3pt}
    \label{fig:addlkernelspow}
\end{subfigure}
\caption{Performance and power results of various kernels. For the Inverse Helmholtz and Interpolation kernels, we also include the performance and estimated efficiency of the corresponding highly-optimized Intel implementations \cite{Rink2018}. All experiments are executed using double-precision floating points.}
\vspace{-6pt}
\end{figure}
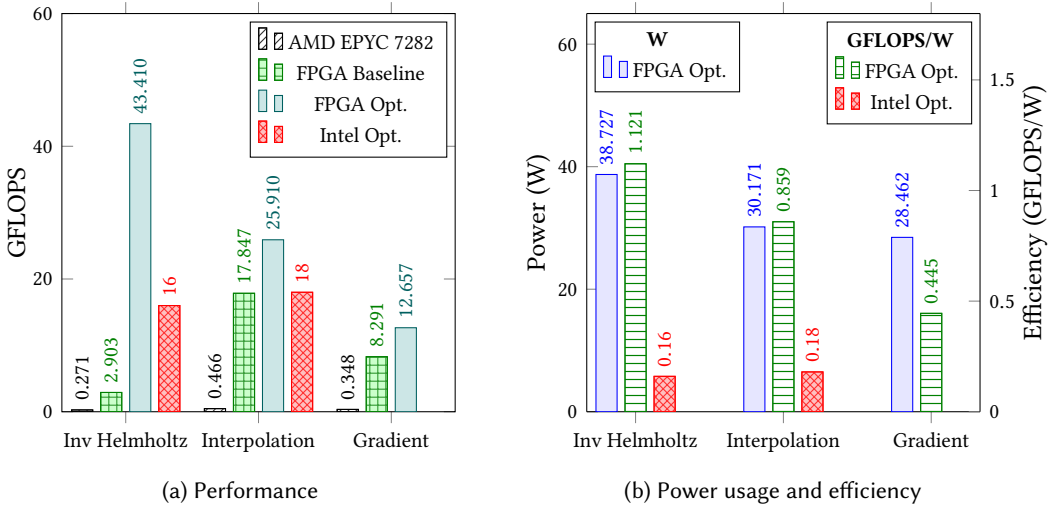

\autoref{fig:addlkernels} shows the performance results for the \textit{Inverse Helmholtz}, \textit{Interpolation}, and \textit{Gradient} kernels on the \amd\ and the FPGA. These results include the AMD execution (black bars), baseline (no optimizations) FPGA implementation (green bars), and the fully-optimized FPGA implementation (azure bars). The fully optimized kernels all use double-precision floating-point data and implement the Double Buffering, Bus Opt (Parallel), and Dataflow (where each loop nest is a subkernel module) optimizations. The baseline implementations achieved $10.7\times$~-~$38.3\times$ speedup over their software execution on the \amd. The optimized FPGA implementations, however, achieved $36.4\times$~-~$160.2\times$ speedup over the AMD execution.

To compare our results with state-of-the-art software implementations, \autoref{fig:addlkernels} also includes the performance of highly-optimized Intel implementations for the \textit{Inverse Helmholtz} and \textit{Interpolation} kernels \cite{Rink2018} (red bars). These implementations are generated by the original CFDlang compiler using the process described in \cite{Rink2018}, which was found to outperform expert-crafted manually optimized kernels. The executables are compiled with the Intel Compiler and use the Math Kernel Library 2017.2.174 and were profiled on a 24-core Intel Xeon E5-2680 v3 CPU (Haswell), running at 2.50~GHz. The FPGA-optimized \textit{Inverse Helmholtz} and \textit{Interpolation} kernels achieved  $2.7\times$ and $1.4\times$ speedup over the optimized Intel execution, respectively.

The average power and efficiency for the optimized kernels are shown in \autoref{fig:addlkernelspow}. The estimated efficiencies (GFLOPS/W) of the Intel executions for the \textit{Inverse Helmholtz} and \textit{Interpolation} kernels are also shown. These estimations are calculated using the GFLOPS results of the kernel and a conservative estimate of the average power (100 W), assuming the CPU would be operating under a lower load than the thermal design power (120 W). The \textit{Interpolation} and \textit{Inverse Helmholtz} kernels are $4.8\times$ and $7.0\times$ more efficient than the Intel CPU execution, respectively. Recalling the results from \autoref{fig:pow_cu1p11}, the most power-efficient implementation of the \textit{Inverse Helmholtz} (32-bit fixed-point with $p=11$ and 1~\gls{cu}) is $24.5\times$ more efficient than the Intel execution.

\section{Concluding Remarks}
Numerical simulations are compute-intensive \gls{hpc} applications used to solve complex problems in many scientific fields. Such applications benefit greatly from parallelization. In this context, HBM FPGA devices are increasingly used to achieve high performance with high energy efficiency. However, designing \gls{hbm} architectures for such systems is complex and requires specific skills. Our analysis of such architectures reveals that the high cost of communication between CPU and FPGA memories and the limited amount of resources to implement parallel kernels are the major issues. To address these challenges, we redesigned a \gls{dsl} compiler in MLIR to automatically generate \gls{hls}-ready code, along with an \gls{hls}-based flow to automate the generation of optimized system architectures that implements several memory-related optimizations. 
Our \gls{mlir} framework offers much quicker turnaround times in implementing the language.
The differences in flexibility between our custom IR and the new dialects were negligible, while diagnostics, stability, and composition are greatly improved.
We have created an opportunity to apply our strategies, even if partially, to other \gls{mlir}-based flows to achieve a more direct comparison with our results in the future.

Our results show that the data format can have a significant impact on performance; a smaller data format simplifies the logic and allows the circuit to have a shorter overall latency and to operate at a higher frequency. The polynomial degree, an application-specific parameter, can also impact performance for similar reasons. These parameters also reduce the total FPGA resources needed to perform the computations, allowing for the possibility of instantiating multiple compute units in the FPGA fabric. However, replication does not equate to increased performance unless the host data transfers can be significantly reduced. If the host data transfers limit the application, the design can be optimized for power efficiency by only instantiating one compute unit. However, if the host were interfaced with multiple FPGAs and were able to send data in parallel to all of them, replicating the compute units onto separate FPGAs would achieve increased performance. 
Overall, we achieved up to 103 GFLOPS--more than 6$\times$ faster than highly-optimized Intel implementations--with an energy efficiency of about 4 GFLOPS/W--almost $24\times$ more efficient than highly-optimized Intel implementations. 

\begin{acks}
This project is partially funded by the EU Horizon 2020 Programme under grant agreement No 957269 (EVEREST).
\end{acks}

\renewcommand{\bibfont}{\footnotesize}
\printbibliography

\end{document}